 \documentclass[final,5p,times,twocolumn]{elsarticle}

\usepackage{mathrsfs}
\usepackage{graphicx}
\usepackage{subcaption}
\usepackage{float}
\usepackage{booktabs}
\usepackage{tabularx}
\journal{.}
\usepackage{xcolor}      
\usepackage{url}
\usepackage{amsfonts}
\usepackage[T1]{fontenc}
\usepackage{amssymb}
\usepackage{amsthm}
\usepackage{latexsym,amsmath}
\usepackage{adjustbox}
\usepackage{etoolbox}
\makeatletter
\def\fps@figure{!htbp}
\def\fps@table{!htbp}
\makeatother

\patchcmd{\section}{\begingroup}{\vspace*{-\baselineskip}\begingroup}{}{}
\patchcmd{\subsection}{\begingroup}{\vspace*{-\baselineskip}\begingroup}{}{}

\newcommand{\softpagebreak}{\ifdim\pagegoal<\maxdimen\pagebreak\fi}

\usepackage{hyperref}

\begin{document}

\begin{frontmatter}



\title{An evaluation of LLMs for political bias in Western media: Israel-Hamas and Ukraine-Russia wars}


\author[inst1,inst2]{Rohitash Chandra \corref{cor1}}
\author[inst1,inst2]{Haoyan Chen}
\author[inst1,inst2]{Yaqing Zhang}
\author[inst1,inst2]{Jiacheng Chen}
\author[inst1,inst2]{Yuting Wu}

\affiliation[inst1]{Transitional Artificial Intelligence Research Group, School of Mathematics and Statistics, UNSW Sydney, Sydney, Australia}
\affiliation[inst2]{Centre for Artificial Intelligence and Innovation, Pingla Institue, Sydney, Australia }
\cortext[cor1]{Corresponding author (e-mail: rohitash.chandra@unsw.edu.au)}



\begin{abstract}
Political bias in media plays a critical role in shaping public opinion, voter behaviour, and broader democratic discourse. Subjective opinions and political bias can be found in media sources, such as newspapers, depending on their funding mechanisms and alliances with political parties. Automating the detection of political biases in media content can limit biases in elections. The impact of large language models (LLMs) in politics and media studies is becoming prominent. In this study, we utilise LLMs to compare the left-wing, right-wing, and neutral political opinions expressed in the Guardian and BBC. We review newspaper reporting that includes significant events such as the Russia-Ukraine war and the Hamas-Israel conflict. We analyse the proportion for each opinion to find the bias under different LLMs, including BERT, Gemini, and DeepSeek. Our results show that after the outbreak of the wars, the political bias of Western media shifts towards the left-wing and each LLM gives a different result. DeepSeek consistently showed a stable Left-leaning tendency, while BERT and Gemini remained closer to the Centre. The BBC and The Guardian showed distinct reporting behaviours across the two conflicts. In the Russia-Ukraine war, both outlets maintained relatively stable positions; however, in the Israel-Hamas conflict, we identified larger political bias shifts, particularly in Guardian coverage, suggesting a more event-driven pattern of reporting bias.
These variations suggest that LLMs are shaped not only by their training data and architecture, but also by underlying worldviews with associated political biases.



\end{abstract}



\begin{keyword}
Political Bias \sep LLMs\sep The Guardian\sep  BBC\sep language models
\end{keyword}

\end{frontmatter}


\section{Introduction}
\label{sec:Introduction}

Mass media, including social media, served as a medium for the dissemination of knowledge that has been used as a medium for political gain and manipulation \cite{dunaway2022mass}. Media immensely contributes to changing citizens' political preferences; for example, more partisan media has contributed to political polarisation and led Americans to support more partisan policies and candidates \cite{prior2013media}. The media's ability to shape public opinion highlights its critical role in shaping political landscapes. For instance, Zinn \cite{zinn2015people} presented a study on mainstream historical narratives, describing the experiences of marginalised groups in American history and emphasising the political nature of historical writing. Thus, text written by individuals inherently carries their own subjectivity and societal experience \cite{gavelek2014ways}. Therefore, regardless of how objective one claims their discourse to be, their upbringing and online environment can subtly influence their perception and judgment. Consequently, the political bias of Western media exists with right, left and centre-leaning political discourse  \cite{mcchesney2016rich,herman2010manufacturing}. Social media platforms enable people to express their opinions freely, but that also comes at a cost, as employees have paid repercussions by losing their jobs for expressing themselves. Statements from users and \textit{influencers}  on social media platforms such as Twitter and Facebook often have political biases, particularly on topics related to elections, free speech, and the role of democracy \cite{himelboim2013birds}, and the political bias extends to news media (newspapers and television).


Traditional methods of detecting political bias in news media involve human identification \cite{guess2019less}, where trained human experts can analyse news articles and identify instances of bias, such as biased political statements that can influence election outcomes. Although this is a time-consuming and labour-intensive approach, it can provide a nuanced and accurate assessment of political bias in news media \cite{gruzd2014investigating}. Despite the potential benefits of these methods, there are limitations in detecting political bias in news media. For instance, Caliskan  \cite {caliskan2017semantics} argued that certain forms of bias may be more subtle and complex to detect through human identification. Additionally, if humans hold politically biased beliefs influenced by their growing environment, it may impact the direction of content analysis \cite{elo2014qualitative}. This influence is evident earlier in the information retrieved by search engines. The algorithmic ranking of search tools can also influence human judgment, as individuals generally tend to accept the viewpoints encountered first and click on the first-ranked link, perceiving it as correct and reliable \cite{pan2007google}. This can lead to preconceived notions and cognitive biases in subsequent information encountered. For example,  Epstein \cite{epstein2015search} derived that biased search rankings can change the voting preferences of undecided voters by more than 20\%. Therefore,there is scope of  NLP and machine learning to know whether the content of news media exhibits political bias and provide potential guidance to readers. Besides, whether a specific journal exhibits changes in political bias over time in the long run, like a comparison between the pre-pandemic era and the post-pandemic era, is also considered.

Natural language processing (NLP) \cite{mihalcea2006nlp} is a subfield of artificial intelligence that focuses on the study of human language using computational tools and machine learning models.  NLP enables computers can understand, create, and even respond to human language \cite{collobert2008unified}, which has gained widespread attention through technologies such as Large Language Models (LLMs) \cite{yao2024survey}. LLMs have emerged as cutting-edge artificial intelligence systems that can process and generate text with coherent communication and generalise to multiple tasks \cite{naveed2023comprehensive}.  Applications of NLP include machine translation, text summarisation, chatbots, sentiment and semantic analysis, interactive user interfaces, multilingual information retrieval, and speech recognition \cite{chowdhary2020natural}.  
ChatGPT \cite{ray2023chatgpt}, Gemini and DeepSeek \cite{liu2024deepseek} are implementations of LLMs that are at the forefront of NLP that enabling users to create highly sophisticated and versatile applications in various tasks and domains, including text generation, translation, and data analysis. 
Studying potential political bias using deep learning and NLP methods is essential because the information individuals encounter can affect their judgment to a certain extent \cite{caliskan2017semantics}.

Political bias is commonly categorised as liberal (left wing), conservative (right wing), and neutral \cite {gezici2022quantifying}; however, the existence of political bias among individuals varies across different countries. 
The left-wing political bias means the group or person holds liberal ideas and hopes to be progressive. The right-wing political bias means the group or person hold the conservative view. For instance, in the United States, traditional classifications include the Republican Party and the Democratic Party \cite{layman2006party}, while in Australia, the Australian Labour Party represents workers' interests. As a result, political bias in different news media outlets can be highly complex. It can be helpful to classify different political parties as left-wing or right-wing. Scholars have conducted extensive research in this area, and several machine learning approaches have been developed to detect political bias in textual content effectively. David et al. \cite{david2016utilizing} attempted to predict the political party affiliation of Israeli Facebook users by machine learning models. Malouf and Mullen \cite{malouf2008taking} used a Naïve Bayes text classifier to predict the political party affiliation of political discussion board users. Kang et al. \cite{kang2022quantifying} used a document classification technique (doc2vec) to train data from Facebook posts to assess political bias in a newspaper in South Korea quantitatively. Chandra et al. \cite{chandra2021biden}  used the BERT language model for Twitter sentiment analysis for the 2020 US presidential elections and reported that user sentiments can be a good proxy for state-wise election outcomes.

    We consider geopolitical events such as the Russia-Ukraine war and the Hamas-Israel conflict, which are widely reported by Western media. During this period,   BBC and the Guardian published a large number of news reports, each of which may contain a specific ideological framework \cite{abdulmehdi2024bais}. The public's understanding of these conflicts, as well as the overall narrative of war, humanitarian issues, and foreign policy, is often influenced by the political perspective adopted in media coverage. According to the results of a field experiment by  Broockman and  Kalla ~\cite{broockman2022crosscutting}, users' political inclinations change according to the news media from which they obtain information.
    
 Therefore, understanding the political tendency of media bias (such as left-wing/right-wing/neutral stance) has a key impact on public cognition, while traditional manual methods for analysing political tendencies (such as manual annotation or keyword analysis) have obvious limitations, such as insufficient scalability and susceptibility to subjective interference.  LLMs  offer significant improvement in the analysis ability of abstract concepts such as political tendency and sentiment detection. A BERT-based classifier achieved 75\% accuracy in political stance classification for YouTube videos ~\cite{aldahoul2024bert}.  Unlike traditional media bias identification methods, LLM has more delicate semantic parsing capabilities, allowing us to systematically detect political bias at a macro scale and time dimension. However, it is worth noting that LLM itself may also imply political stance bias caused by training data~\cite{peters2022bias} ~\cite{kumar2024llmbias}. Different LLM model architectures may produce different results for bias identification of the same text, and this "detector bias" may affect the objectivity of the analysis conclusion.

  Previous studies on \textit{detecting political bias using LLMs} have mostly focused on a single event or a single model, and rarely considered a comprehensive analysis of multiple media, multiple models, and different periods.

    Political bias in the media is not only a matter of editorial choice; it is also shaped by context, such as global events and the tools used for analysis and interpretation. In this study, our objective is to explore three dimensions of political bias in Western media coverage:

    {\bigskip\noindent} 1. How does political bias in the same media outlet (e.g. BBC or The Guardian) change before and after major geopolitical events, such as the Russia–Ukraine war?

    {\bigskip\noindent} 2. Are there observable differences in how different media outlets present political perspectives in a given timeframe when reporting on the same event?


\par\bigskip
In this paper, we present a study that employs LLMs for evaluating political bias in selected news media covering significant events, including the Russia-Ukraine war and the Hamas-Israel war. We select the Guardian newspaper and BBC news for analysis and provide a comparison of LLMs that include Gemini and DeepSeek to review political leaning (left and right wing) over specified periods. The objective of this project is not only to detect political bias in media, but also to understand how language models and media institutions differ in their representation of political events. We apply a combination of  NLP and LLMs to systematically analyse a large-scale news dataset.  We design a timeline-based analysis framework to capture shifts in media bias before and after major geopolitical events and use multiple LLMs (including BERT,  DeepSeek, Gemini) to classify articles by political leaning and compare model behaviours. 


\section{Related Work}

\subsection{Political leaning in news media}

The study of political bias in the media has attracted significant attention, as it has shaped political narratives that have led to conflicts and wars. For example, the left and right-leaning newspapers’ long-established beliefs and agenda motivated readers to voice their opinions about the Kyle Rittenhouse case \cite{perez2025comparative}. Traditionally, the political biases are identified based on respondents’ input, typically gathered through surveys and questionnaires and established methods of investigation \cite{guess2019less}. An example of such a method is whether the news article favours voting for Republicans or Democrats in the 2008 federal elections \cite{zhou2011classifying} using a keyword analysis. Furthermore, statistical models such as the Random Effects model have been utilised to make judgments where the key independent variable (political leaning) reflects which contents will contribute to the political bias \cite{shor2019political}. However, the result may not be so accurate when dealing with a complicated article, which can be improved by machine learning. Jiang et al.  \cite{jiang2008finding} developed a political vs. non-political blog classifier using a bag-of-words model and SVM (Support Vector Machines) and found high accuracy by bag-of-words features. 

\subsection{LLMs for analysis of political leaning}

Conventional machine learning  and NLP models built the foundations for detecting the political leaning in news media. However, they often faced challenges in scaling, handling multilingual
 and unstructured data, and deriving insights from vast corpora of text \cite{li2024political}. The emergence of the LLMs overcomes the shortcomings of the traditional method, providing the researchers with a tool for analysing political bias with enabling an automated, large-scale technique. Recent studies used the GPT-4 model to judge the political bias of an article, where they found the accuracy of GPT-4 was  outperforming conventional  models \cite{heseltine2024large}.   Furthermore, the study used different LLMs to investigate how advancements in these models impact their performance and outputs, where ChatGPT-4 is found better than other models \cite{heseltine2024large}.

\subsection{Sentiment analysis of news and social media}

The application of sentiment analysis using  language models has gained significant attention in recent years. We can know the sentiment of an article during the period of some global crisis, such as news media discussion and trends during the COVID-19 pandemic \cite{chandra2025large}. For example, during the COVID-19 crisis, researchers have largely focused on social media platforms as sources for real-time sentiment extraction. Chandra et al.  \cite{chandra2021covid} analysed tweets (X) from India and found a rich variety of sentiments, including combinations of optimism, humour, and anxiety as the number of deaths and infections rose in the second phase of COVID-19. Chandra and Saini \cite{chandra2021biden} demonstrated that sentiment analysis can be useful in identifying voter trends, taking the United States 2020 Presidential Elections as a case study. Furthermore, Chandra et al. \cite{chandra2024analysis} demonstrated that sentiment analysis can be used as a tool for analysis of anti-vaxer sentiments during COVID-19. Lande et al. \cite{lande2023deep} used topic modelling for analysis of tweets during COVID-19 and mapped it to major events for major COVID-19 variants.

\section{Methodology}

\subsection{Data}
    


     We constructed four datasets from two media sources, including BBC and The Guardian. We cover two geopolitical conflicts, spanning key periods of the  Russia-Ukraine war and the Hamas-Israel conflict \cite{yabbies2025guardiandata} \cite{yabbies2025bbcdata}. We chose a five-year period (January 2020 to December 2024) that captures a reasonable length of the pre-war and during-war phases of each conflict. We selected this five-year temporal window for two main reasons: (1) to ensure comparability across datasets by maintaining a uniform time frame, and (2) to allow sufficient time before and after each war and capture potential shifts in patterns. 

 Given that the BBC does not provide a public API (Application Programmer Interface) for automated retrieval of news articles, we utilised the {RealTimeData/bbc\_news\_alltime} dataset available on Huggingface \cite{realtimedata2023bbc}, which provides article web links, publication dates and titles spanning from January 2017 to 2025; note that our study screened the captured data from 2020 to 2024. However, this dataset does not include the body contents of the articles, which is essential for our LLM-based political bias analysis. Therefore, we developed a web scraper using Python (available in our Github repository \footnote{\url{https://github.com/sydney-machine-learning/politicalbias-LLMs/codes}}) to retrieve the full articles using the web links. 
    
  We applied keyword-based filters on the article contents during scraping to focus our analysis on relevant geopolitical contexts.  In the case of the Russia–Ukraine war, we selected keywords "Russia" and "Ukraine" to ensure a broad coverage for both wartime articles and pre-war reporting. We followed a similar approach for the Hamas-Israel conflict; we used the broader keyword set "Israel", "Hamas", and "Palestine" to capture the wider context within which the October 2023 war later erupted. We retained articles matching these keywords with their extracted full texts.   We accessed the Guardian data via their  API using a  Python-based scraper and matching the same set of keywords used for the BBC. We restricted our search to articles published between 2020 to 2024 and retrieved records that include title, content, publication date, and web links, available in our Github repository. We analysed the length of content for each record and found that the majority of articles (above 95 percent) have less than 10,000 tokens. Hence, we removed the small margin of articles that were longer than 10,000 tokens for better time-effectiveness and cost-efficiency. Furthermore, very long texts may contain invalid content or repetitive structures, which may weaken the focus of the political framework. Removing these values can ensure that delay, cost, and noise in bias analysis are reduced. The datasets are available our GitHub repository \footnote{\url{https://github.com/sydney-machine-learning/politicalbias-LLMs/data}}, a summary of the datasets is in the table below. 

\begin{table*}[h]
\centering
\small
\begin{tabular}{|l|l|r|l|r|}
\hline
\textbf{Dataset} & \textbf{Source} & \textbf{Article Count} & \textbf{Date Range} & \textbf{Size} \\
\hline
BBC-Hamas-Israel       & HuggingFace      & 1829  & 01/2020–12/2024 & 9.3 MB \\
BBC-Russia-Ukraine     & HuggingFace      & 3912  & 01/2020–12/2024 & 18.5 MB \\
Guardian-Hamas-Israel  & Guardian API & 5323   & 01/2020–12/2024 & 37.9 MB \\
Guardian-Russia-Ukraine& Guardian API & 11002  & 01/2020–12/2024 & 77.5 MB \\
\hline
\end{tabular}
\caption{Overview of datasets used in this study}
\label{tab:dataset-overview}
\end{table*}

\subsection{N-gram}
Using the n-gram model, we can extract classes that have the flavour of either syntactically based groupings or semantically based groupings \cite{brown1992class}. The model n-gram is a contiguous sequence of n items and is based on the frequency of these times to predict upcoming words from a given sequence of text \cite{nagalavi2016n}. In this study, we use n-gram analysis  for analysis of topical relevance of selected segments of the news data.

\subsection{LLMs}
Although the  BERT \cite{devlin2019bert}   is not a generative model, it serves as a foundational predecessor to a series of subsequent LLM models. BERT was  designed as a pre-trained model for NLP tasks and is prominently used for word embedding for classification tasks such as sentiment analysis \cite{hoang2019aspect}, and topic classification \cite{shah2023multitask,li2025englishtopic}. In this study, we employed a pre-trained BERT model \footnote{\url{https://huggingface.co/bucketresearch/politicalBiasBERT}} designed explicitly for political bias classification to determine the political leanings of the Western media-sourced articles without any additional fine-tuning. Although  BERT inherently restricts input to 512 tokens, simply truncating the text can result in information loss and disrupted context, which may affect the accuracy of classification. Therefore, efficiently and comprehensively handling overlength documents is considered one of the key challenges in this study. Therefore, we use a sliding window chunking combined with a majority voting mechanism, where we tokenise the input text  into a sequence $T = \{t_1, t_2, \ldots, t_n\}$. We then apply a sliding window with size $w = 512$ and stride $s = 256$  to generate overlapping chunks until the entire text is covered. Then, the final prediction is determined by majority vote among the chunk-level predictions.


 LLMs have been undergoing rapid evolution to address the demanding needs of different fields in academia and the industry. Gemini 1.5 Flash, released by Google DeepMind in May 2024 ~\cite{gemini2024deepmind}, represents a new generation of LLMs developed for high-throughput and long-context tasks. It adopts the Mixture-of-Experts (MoE) technique, which selectively activates expert modules during analysis; hence, reducing computational cost without compromising performance. A key feature of Gemini 1.5 Flash is its high tolerance to long contexts, i.e. up to 1 million tokens, which allows the model to process entire articles holistically, removing the need for truncation or sliding window chunking ~\cite{geminiapi2024docs}. This feature makes it particularly suitable for detecting political bias from long-context articles in our study. While more recent models such as Gemini 2.0 and 2.5 offer enhanced multi-modal reasoning and broader context support ~\cite{gemini2023blog,gemini2024googlecloud}, Gemini 1.5 Flash was selected for this study for its accessibility, time-efficiency and cost-effectiveness. In our study, Gemini 1.5 Flash was accessed via Gemini's official API using a prompt technique.

 DeepSeek-V3 ~\cite{liu2024deepseek} also adopts the MoE technique with a total of 67.1 billion parameters, of which only 3.7 billion are activated per token ~\cite{deepseek2024}. This design enables low-cost and inference efficiency for large-scale analysis. In contrast to many "Western" LLMs (such as ChatGPT, Gemini, Claude), DeepSeek incorporates self-distillation techniques, leveraging self-generated data to reduce dependence on external sources~\cite{ibm2025, eden2025}. These characteristics make DeepSeek-V3 particularly valuable for exploring cross-cultural interpretations of political discourse and for evaluating model behaviours between Chinese and Western LLMs.

 Unlike the BERT-based classifier, Gemini and DeepSeek models are prompted to give feedback for each article and thus also play the role of classifiers. We constructed a single prompt to each full-length article to ensure consistency across Gemini and DeepSeek models, prompting the model to output exactly one of three labels: ``Left'', ``Right'', or ``Centre''. Note that we do not provide contextual constraints or definitions for what "Left", "Right" or "Centre" each represents; this allows the models to interpret political leaning based on their own learned representation and vocabulary. The same prompt structure and temperature setting ($T=0.0$) are used to avoid randomness and constrain response length~\cite{brown2020language, hinton2015distilling}. We process each article multiple times to improve reliability,  and select the most frequent label as the final output. We fix all other parameters throughout the runs, such as input content, stop tokens, and token limits. We adopt multi-threaded processing to support large-scale computation, and API token rate limits are overcome by retry and slow-down logic.

\subsection{Sentiment Analysis}


 Although the classification of political bias using LLMs remains the primary focus of this study, sentiment analysis is complementary to this primary focus. Sentiment analysis provides an alternative perspective for examining the intertwining relationship between emotions and political bias in media. RoBERTa is a pre-trained transformer model that has demonstrated solid performance in sentiment classification, improving from BERT by incorporating dynamic word masking during training~\cite{liu2019roberta}. In this study, we adopted a fine-tuned DistilRoBERTa model\footnote{\url{https://huggingface.co/j-hartmann/emotion-english-distilroberta-base}}, which is specifically trained for emotion classification in English; as the name suggests, DistilRoBERTa model is a distilled version of RoBERTa~\cite{hartmann2023emotion}, reducing parameters from 125 million to 82 million. The distilled version reduces computational complexity while improving speed and maintaining performance~\cite{sanh2019distilbert}, making it suitable for large-scale classification tasks. The DistilRoBERTa predicts emotions in Ekman emotion dimensions, including joy, anger, fear, sadness, neutral, disgust and surprise~\cite{hartmann2023emotion}.  

 In this study, we apply the DistilRoBERTa model to full articles in each dataset. The predicted emotion probabilities were aggregated to produce summaries and visualisations of the emotional shifts before the breakout and during the course of each war. This method allows us to observe how emotions change over time and explore how patterns might correlate with trends in political leanings.

\subsection{Framework}
\begin{figure*}[t]
    \centering
    \includegraphics[width=\linewidth]{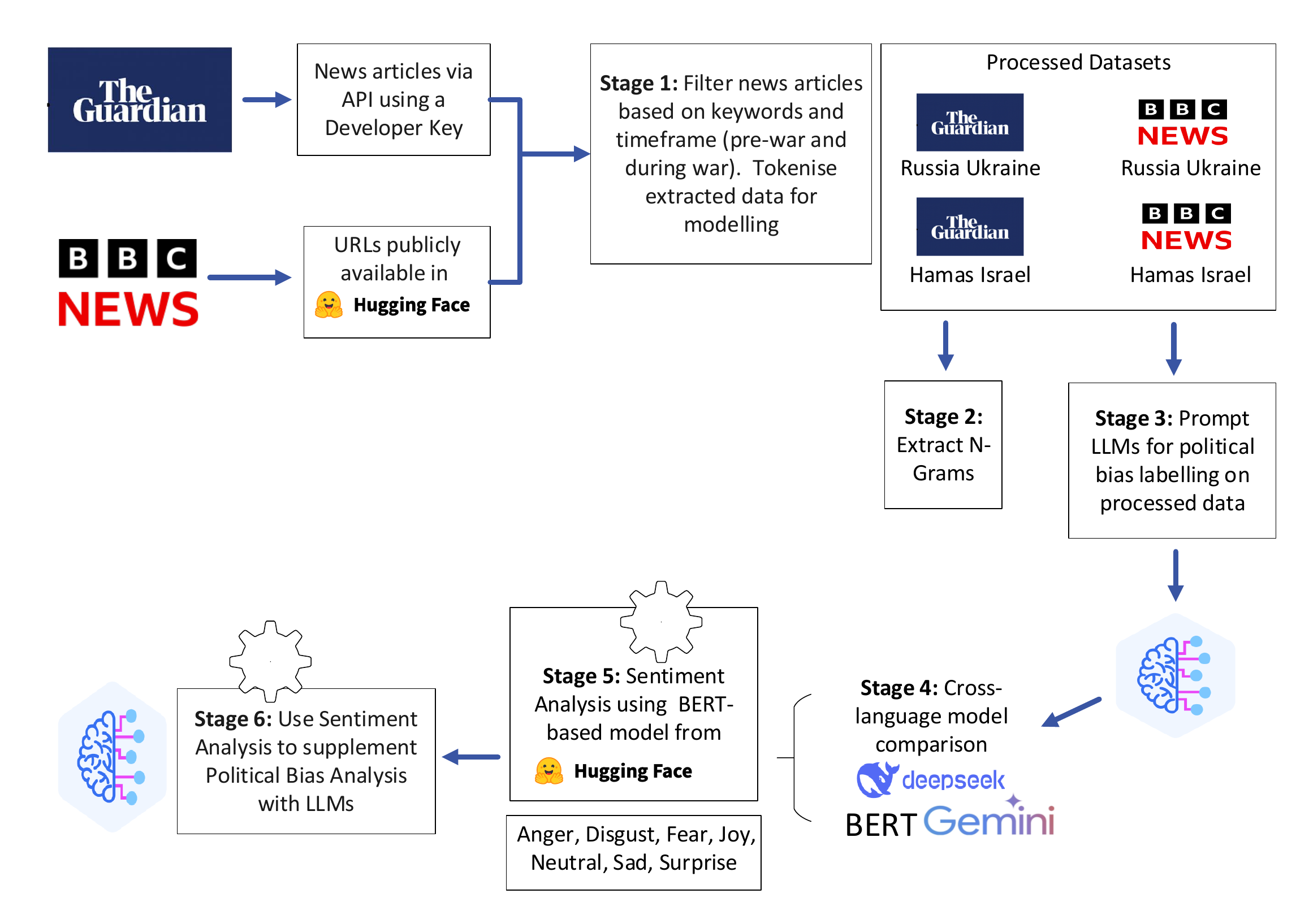}
    \captionsetup{justification=centering}
    \caption{Overall framework of analysis of political biases in Western media for this study.}
    \label{fig:workflow}
\end{figure*}



We developed this analytical framework featuring data processing, political bias classification using LLMs, and finally sentiment analysis as shown in  Figure \ref{fig:workflow}.

Step 1 involves the extraction (scraping) of data from the Guardian and BBC using their API and developer keys \footnote{\url{https://open-platform.theguardian.com}}.   
We also define the keyword sets used for filtering, as the right set of keywords is crucial for this step and the rest of the study. For instance, using “war” as a keyword would most likely collect articles explicitly containing “war”. However, before the outbreak of a war, such terms might not appear in relevant articles. Since our study aims to compare political bias and sentiments before and after the outbreak of war, we need keywords that are not only highly relevant to the intended context but also consistently applicable across both prewar and wartime periods. After a rigorous search and exploration, our solution to this dilemma was to extend the underlying context from “war” to the political entities involved.  Clausewitz \cite{clausewitz1832onwar} said that war is a political act carried out by other means. Therefore, our keywords for the Russia–Ukraine conflict were simply \{“Russia”, “Ukraine”\}, while those for the Hamas–Israel conflict included \{“Hamas”, “Palestine”, “Israel”\}.   
We leveraged the official Open Platform API~\cite{guardianapi} to collect Russia-Ukraine related articles from The Guardian, which offers structured metadata for each article (publication date, section, URL, title, contents, and editorial tags). Instead of content filtering, we filtered article topic tags "world/russia" and "world/ukraine", which allowed us to retrieve articles associated with each geopolitical entities. Similarly, for Hamas-Israel related articles, we filtered article topic tags "world/hamas", "world/israel" and "world/palestinian-territories". It is worth noting that "palestinian-territories" is not the same as the chosen keyword: "Palestine", because The Guardian does not have a "Palestine" tag, we choose its semantically close substitute - "palastinian-territories'. 

 We then scraped articles from the BBC using publicly available URLs from Hugging Face. The bbc\_news\_alltime dataset on Hugging Face\footnote{\url{https://huggingface.co/datasets/RealTimeData/bbc_news_alltime}} does not include full texts of the news articles, but the URLs has made the scraping of news contents possible. Hence, we developed another web scraper with Python that uses the URLs to collect the full articles \footnote{Web scraper: \url{https://github.com/sydney-machine-learning/politicalbias-LLMs/codes}}. These articles do not have editorial tags, so we filtered the contents using keywords in the process of web scraping, retaining only the articles that contained the keywords. To be consistent with The Guardian datasets, we used \{“Russia”, “Ukraine”\} as keywords for Russia-Ukraine, and \{“Hamas”, “Palestine”, “Israel”\} for Hamas-Israel. The datasets also included publication date, URL, title, and contents.

We apply standard data cleaning techniques to all four datasets (Table \ref{tab:dataset-overview}), including removal of null values,  special characters, non-English words, and removal of encoding errors. We used URLs as the unique identifier for the articles and checked for any duplication based on URLs. 

We also feature additional data preparation to optimise cost-effectiveness and time efficiency for the data processing and analysis by the respective LLMs. We tokenise the articles and analyse the distribution of token count, and the vast majority of articles had a token count of less than 10,000, with less than 3\% of data having a token count of over 10,000. Given each political bias label account for one vote in the final poll regardless of its length, and the articles with high token count requires more computational resources, yet the LLMs have an upper limit on number of tokens they each could process; hence we removed articles with more than 10,000 tokens to ensure uniform model performance and avoid incomplete articles processing, which can affect political leaning or sentiment inference. At this stage, the four datasets were prepared, with each rows containing URL, title, contents, and publication date. Specifically, the publication date is in the same date format, which enables time-series analysis and dataset partitioning into "pre-war" and "during-war" subsets.

Step 2 features N-gram analysis to assess the datasets on their relevance to the underlying topics. N-gram analysis not only allows us to identify high-frequency words and key linguistic patterns but also visually confirms the relevance of the datasets to the intended topics. We removed stop words for n-gram analysis using the SpaCy \footnote{\url{https://spacy.io}} package in Python. We also customised a set of stop-words containing common but semantically neutral words found in our articles, such as "say", "state", "old", "tell", "bst", "gmt", etc. We then undertake   Bigram and Trigram for each dataset to review the topic relevance of our datasets.

Step 3 involves political bias classification using the respective models for every article collected in our datasets with labels including Left, Centre, or Right. We utilise LLMs including: BERT from Hugging Face (pre-trained for political context), Gemini 1.5 Flash, and DeepSeek-V3. In the case of BERT, the initial challenge was the model's token limit of 512, resulting in articles being truncated. We implemented a novel approach, which involves sliding window chunks with a voting mechanism. We sectioned (bagged) the article into overlapping paragraphs of 512 tokens, each processed individually by the model for inference. Followed by a voting sequence to determine the article-level political leaning based on the majority labels across paragraphs. For Gemini and DeepSeek, we followed a uniform approach to maintain comparability.  Each article was processed in full (with a 10,000-token limit) and prompted to assign a label (Left, Centre, or Right) to the article. When developing the LLM Prompt, we intentionally avoided defining "Left", "Right", or "Centre" within the prompt to prevent imposing external constraints on the model's interpretation. This allows the LLM to apply its definition of political leaning when classifying the articles. 


In Step 4, we analyse the respective LLMs' classification individually. From the last step (step 3), we received datasets with political bias labels, with the ability to be further segregated into "pre-war" and "during-war" subsets by dates. We generated longitudinal charts comparing the changes of political bias in time series (pre-war vs. during-war) and across events (Russia-Ukraine vs. Hamas-Israel).
In this step, we conducted a comparative analysis of BERT, Gemini, and DeepSeek. It is acknowledged that political leaning is indeed a subjective matter, and a "true label" may not exist. We argue that a meaningful evaluation of LLMs' political bias classification can only be achieved through comparative analysis across models. The analysis in this step involved longitudinal and cross-sectional model comparisons.

In Step 5, we execute sentiment analysis with DistilRoBERTa \footnote{\url{https://huggingface.co/j-hartmann/emotion-english-distilroberta-base}} to assess the relationship between detected political bias and emotions. The model classifies English texts into one of seven categories: anger, disgust, fear, joy, sadness, surprise, and neutral. It was selected for its ability to handle news reporting articles that tend to use more formal language. Any model that carries a RoBERTa legacy, the DistilRoBERTa model also has a default token count limit of 512\cite{liu2019roberta}. Therefore, we also implemented a sliding window with a voting mechanism, similar to the innovative approach we took to overcome the token count restriction for the political BERT model.

Finally, in Step 6, we aggregate the sentiment analysis results and visualise emotion distributions across media outlets and time series. This allowed us to study the changes in emotions — for example, whether fear or anger increased during wartime coverage — and relate these to observed political leanings. Emotional trends were also compared between the two conflicts.  We gather insights from both political leaning and sentiment analysis.

\subsection{Technical - Implementation Details}

We implemented the framework using Python~3.9 and executed on a workstation equipped with an Intel~Core~i9--13980HX processor, 32~GB of RAM, and an NVIDIA~RTX~4090 GPU. We employed three LLMs for political bias classification, including  Political-BERT\footnote{\url{https://huggingface.co/bucketresearch/politicalBiasBERT}}, DeepSeek-V3\footnote{\url{https://api.deepseek.com}}, and Gemini\footnote{\url{https://ai.google.dev/gemini-api/docs?hl=zh-cn}}.

We implemented sentiment analysis using  \texttt{emotion-english-distilroberta-base} model\footnote{\url{https://huggingface.co/j-hartmann/emotion-english-distilroberta-base}}, which is a fine-tuned version of DistilRoBERTa. This model has been trained using six diverse datasets sourced from Twitter, Reddit, student self-reports, and television dialogues. It is capable of predicting seven emotion classes, including Ekman’s six basic emotions (anger, disgust, fear, joy, sadness, and surprise) as well as a neutral class, with nearly 20{,}000 total training samples. In addition, we used a trigram ($n = 3$) model for textual feature extraction.

\section{Results}
\subsection{LLM results }
\begin{table*}[htb]
\centering
\renewcommand{\arraystretch}{1.2}
\caption{Guardian Articles with Predicted Political Bias Labels by Different Models}
\label{tab:guardian_bias}
\begin{tabular}{p{7.3cm}p{4cm}p{1.1cm}cccc}
\toprule
URL & Title & Published Date & Bert & DeepSeek & Gemini \\
\midrule
\url{https://www.theguardian.com/artanddesign/2022/dec/23/evgeny-maloletka-agency-photographer-of-2022-ukraine} & Ukraine's Evgeny Maloletka: agency photographer of 2022 & 2022/12/23 & Left & Center & Center \\
\url{https://www.theguardian.com/artanddesign/2022/feb/10/russian-painting-vandalised-by-bored-gallery-guard-who-drew-eyes-on-it} & Russian painting vandalised by `bored' gallery guard who drew eyes on it & 2022/2/10 & Right & Center & Center \\
\url{https://www.theguardian.com/artanddesign/2022/feb/23/brave-woman-symbolises-ukraine-mark-nevilles-best-photograph-kremlin-falsehoods} & The brave woman who symbolises Ukraine: Mark Neville's best photograph & 2022/2/23 & Left & Left & Left \\
\url{https://www.theguardian.com/artanddesign/2022/jul/07/russias-sami-fight-to-save-their-language-and-traditions-photo-essay} & Russia's Sami fight to save their language and traditions – photo essay & 2022/7/7 & Left & Left & Center \\
\url{https://www.theguardian.com/artanddesign/2022/jul/20/russian-sami-smokes-fish-natalya-saprunovas-best-photograph-reindeer-sacred-animal} & A Russian Sami smokes some fish: Natalya Saprunova's best photograph & 2022/7/20 & Left & Left & Center \\
\url{https://www.theguardian.com/artanddesign/2022/jun/10/finland-and-russia-a-photo-journey-through-the-border-zone} & Finland and Russia: a photo journey through the border zone & 2022/6/10 & Center & Right & Center \\
\bottomrule
\end{tabular}
\end{table*}

As shown in Table~\ref{tab:guardian_bias}.We consolidated the bias prediction results from four LLM models into four tables, corresponding to datasets from BBC's coverage of the Hamas-Israel conflict, BBC's coverage of the Russia-Ukraine war, Guardian's coverage of the Hamas-Israel conflict, and Guardian's coverage of the Russia-Ukraine war. Taking Guardian's Russia-Ukraine war analysis dataset as an example, it contains article URLs, titles, publication dates, and analysis results from the four LLM models. We will subsequently use the URL as a unique primary key identifier, with each URL representing a distinct article for subsequent analyses.
\subsection{N-gram}

We conducted comprehensive n-gram analyses (bigrams and trigrams) across all four datasets to examine lexical patterns and topic relevance in our media corpus. Although bigrams capture the most frequently occurring two-word associations, trigrams reveal deeper contextual combinations that may reflect recurring discourse structures or implicit political framing. 
We present only the trigram results that directly compare \textit{BBC} and \textit{The Guardian} for each conflict: Russia--Ukraine (Table~\ref{tab:ru_trigrams}) and Israel--Hamas (Table~\ref{tab:ih_trigrams}). 
These grouped bar charts display the most frequent three-word phrases jointly used in both outlets, illustrating vocabulary overlap and divergence in reporting narratives. 
We provide the complete n-gram outputs in our GitHub repository\footnote{\url{https://github.com/sydney-machine-learning/politicalbias-LLMs/tree/main/result/n-gram}}.

\begin{table*}[!t]
    \centering
    \captionsetup{justification=centering}
    \caption{Top 20 Trigrams -- Israel--Hamas: Comparison between BBC and \textit{The Guardian} (with frequency per 10,000 (k) articles)}
    \label{tab:ih_trigrams}
    \vspace{0.5em}
    \begin{adjustbox}{center,max width=\textwidth-1cm}
    \begin{tabular}{c p{0.22\textwidth} c c p{0.22\textwidth} c c}
        \toprule
        \textbf{No.} & \textbf{BBC Trigram} & \textbf{BBC /10k} & \textbf{BBC Count} & \textbf{The Guardian Trigram} & \textbf{Guardian /10k} & \textbf{Guardian Count} \\
        \midrule
        1  & hamas-run health ministry & 6.9 & 1053 & prime minister benjamin & 5.4 & 1827 \\
        2  & prime minister benjamin & 5.9 & 903 & minister benjamin netanyahu & 5.4 & 1823 \\
        3  & minister benjamin netanyahu & 5.8 & 897 & israeli prime minister & 4.2 & 1433 \\
        4  & israel defense forces & 4.6 & 712 & occupied west bank & 3.0 & 1021 \\
        5  & defense forces idf & 3.9 & 597 & israel defense forces & 2.5 & 853 \\
        6  & israeli prime minister & 3.6 & 548 & health ministry said & 2.2 & 760 \\
        7  & president joe biden & 3.3 & 508 & gaza health ministry & 2.2 & 742 \\
        8  & occupied west bank & 2.7 & 411 & un security council & 2.0 & 674 \\
        9  & israeli military said & 2.6 & 402 & israeli military said & 1.9 & 658 \\
        10 & southern israel october & 2.5 & 386 & us secretary state & 1.9 & 631 \\
        11 & people killed gaza & 2.4 & 365 & international humanitarian law & 1.8 & 627 \\
        12 & 200 people killed & 2.3 & 355 & president joe biden & 1.8 & 620 \\
        13 & israel prime minister & 2.3 & 352 & pic twitter com & 1.7 & 582 \\
        14 & killed gaza since & 2.2 & 331 & israel prime minister & 1.7 & 578 \\
        15 & israeli air strike & 2.1 & 322 & state antony blinken & 1.6 & 553 \\
        16 & health ministry says & 2.1 & 318 & secretary state antony & 1.6 & 552 \\
        17 & us president joe & 2.1 & 317 & minister yoav gallant & 1.5 & 497 \\
        18 & times israel reported & 1.9 & 293 & defense forces idf & 1.5 & 495 \\
        19 & israeli air strikes & 1.9 & 287 & international court justice & 1.4 & 487 \\
        20 & bbc co uk & 1.8 & 275 & us president joe & 1.4 & 483 \\
        \bottomrule
    \end{tabular}
    \end{adjustbox}
\end{table*}
\begin{table*}[!t]
    \centering
    \captionsetup{justification=centering}
    \caption{Top 20 Trigrams -- Russia--Ukraine: Comparison between BBC and \textit{The Guardian} (with frequency per 10k articles)}
    \label{tab:ru_trigrams}
    \vspace{0.6em}
    \begin{adjustbox}{center,max width=\textwidth-1cm}
    \begin{tabular}{c p{0.22\textwidth} c c p{0.22\textwidth} c c}
        \toprule
        \textbf{No.} & \textbf{BBC Trigram} & \textbf{BBC /10k} & \textbf{BBC Count} & \textbf{The Guardian Trigram} & \textbf{Guardian /10k} & \textbf{Guardian Count} \\
        \midrule
        1  & president vladimir putin & 6.7 & 1777 & president volodymyr zelenskiy & 5.6 & 3930 \\
        2  & president volodymyr zelensky & 5.7 & 1518 & president vladimir putin & 3.4 & 2386 \\
        3  & russian president vladimir & 3.7 & 993 & pic twitter com & 3.1 & 2199 \\
        4  & russia invasion ukraine & 3.6 & 966 & ukrainian president volodymyr & 2.4 & 1664 \\
        5  & president joe biden & 2.8 & 754 & russia invasion ukraine & 2.3 & 1649 \\
        6  & ukrainian president volodymyr & 2.5 & 655 & defence ministry said & 2.1 & 1471 \\
        7  & us president joe & 2.0 & 545 & ukraine president volodymyr & 2.1 & 1458 \\
        8  & responsible external sites & 2.0 & 529 & russian president vladimir & 2.1 & 1453 \\
        9  & bbc responsible external & 2.0 & 529 & volodymyr zelenskiy said & 1.9 & 1322 \\
        10 & full-scale invasion ukraine & 1.8 & 492 & russia defence ministry & 1.6 & 1097 \\
        11 & world war two & 1.6 & 424 & second world war & 1.5 & 1049 \\
        12 & volodymyr zelensky said & 1.5 & 411 & von der leyen & 1.5 & 1044 \\
        13 & bbc co uk & 1.5 & 402 & president joe biden & 1.4 & 995 \\
        14 & mr zelensky said & 1.5 & 391 & nuclear power plant & 1.3 & 937 \\
        15 & ukraine president volodymyr & 1.4 & 373 & uk ministry defence & 1.3 & 932 \\
        16 & russia full-scale invasion & 1.3 & 343 & russia war ukraine & 1.3 & 917 \\
        17 & co uk please & 1.3 & 337 & foreign minister sergei & 1.2 & 808 \\
        18 & uk please include & 1.3 & 335 & air defence systems & 1.1 & 805 \\
        19 & special military operation & 1.2 & 326 & chancellor olaf scholz & 1.1 & 789 \\
        20 & haveyoursay bbc co & 1.2 & 322 & spokesperson dmitry peskov & 1.1 & 771 \\
        \bottomrule
    \end{tabular}
    \end{adjustbox}
\end{table*}

The trigram  from the Russia-Ukraine context (Table~\ref{tab:ih_trigrams},Table~\ref{tab:ru_trigrams}), we observe that the high-frequency keywords in each set (e.g., \textit{"president vladimir putin", "president volodymyr zelensky", "russia invasion ukraine"}) which are highly relevant to the underlying topic. Similarly, the tri-gram figures from the Israel-Hamas datasets (Table~\ref{tab:ih_trigrams}, Table~\ref{tab:ru_trigrams}) demonstrated the same pattern, where keywords like \textit{"hamas run health", "prime minister benjamin", "israeli defense forces"} are among the highest-frequency groups. The results validate the topic-relevance of our datasets and demonstrate the success of our keyword-filtered data collection method. These terms are highly relevant to the underlying geopolitical contexts of interest - Hamas-Israel and Russia-Ukraine.

Secondly, as part of the exploratory data analysis process, the tri-gram patterns reveal clear perspective differences between \textit{BBC} and \textit{The Guardian} in covering the Russia–Ukraine war. BBC’s high-frequency phrases such as \textit{“president vladimir putin”}, \textit{“president volodymyr zelensky”}, and \textit{“russia invasion ukraine”} shows political leadership and a narrative focused on invasion. Also, the Russian president seems to have been mentioned more than the Ukrainian president by \textit{BBC}. In contrast, \textit{The Guardian} more frequently features the Ukrainian president \textit{“president vologymyr zelenskiy”} over the Russian counterpart \textit{“president vladimir putin”}, and employs more varied terminology to describe the conflict, such as \textit{“nuclear power plant”} and \textit{“second world war”}, reflecting a stronger leaning towards the Ukrainian perspective and multi-angle interpretation. From the Israel-Hamas context, the keywords \textit{"hamas run health"} and \textit{"run health ministry"} made it to the top of \textit{BBC's} list, suggesting a tendency of news reporting from Hamas's perspective. On the contrary, \textit{The Guardian's} tri-gram has \textit{"prime minister benjamin"} and \textit{"minister benjamin netanyahu"} appeared as the highest frequency keywords.

Neutral and steady was the BBC’s reporting before and after the war.  
Though initially left-leaning, more neutral became Guardian's tone later.  
Gemini’s classification proved stable, restrained, and trustworthy.


\subsection{Political leaning  analysis}
\begin{figure*}[!b]
  \centering
  \includegraphics[width=\textwidth]{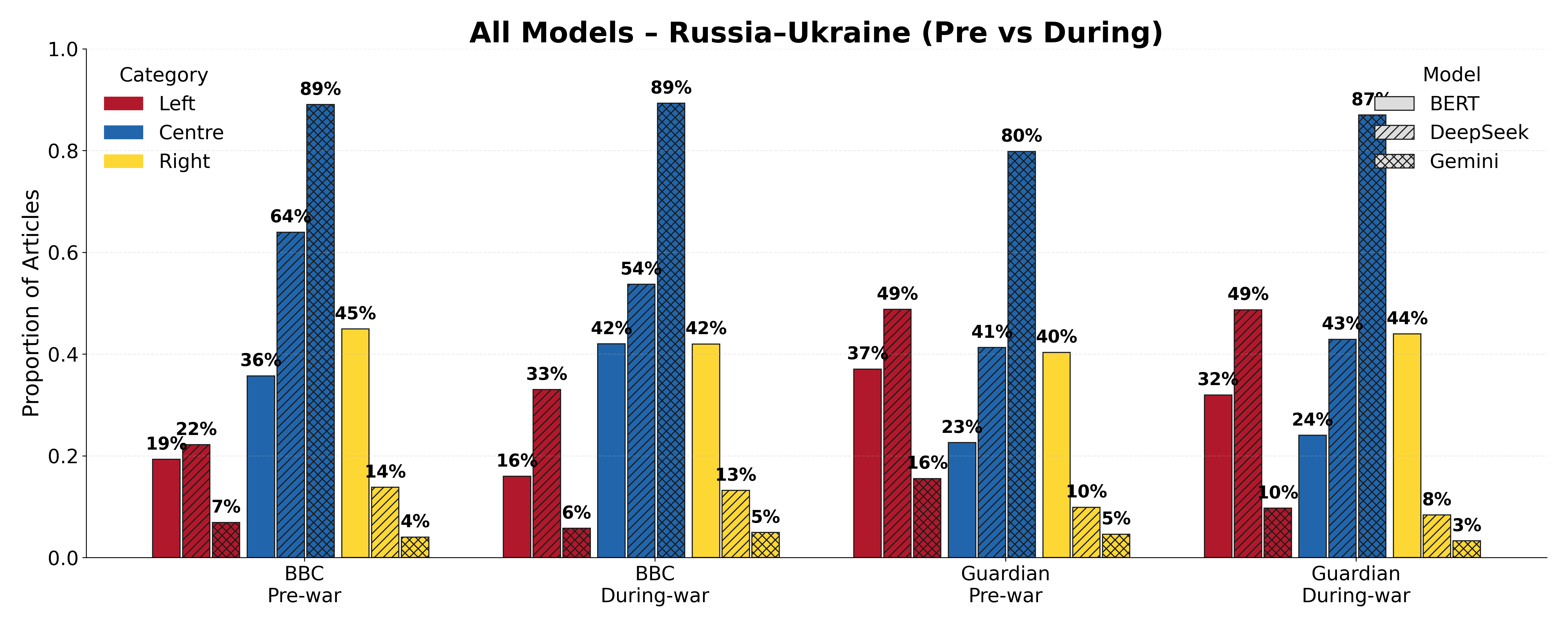}
  \caption{All Models – Russia–Ukraine (Pre vs During). 
  Proportion of articles by Category (Left/Centre/Right) for BBC and \textit{The Guardian};
  hatch fills distinguish models (BERT, DeepSeek, Gemini).}
  \label{fig:all_models_ru}
\end{figure*}

\begin{figure*}[!b]
  \centering
  \includegraphics[width=\textwidth]{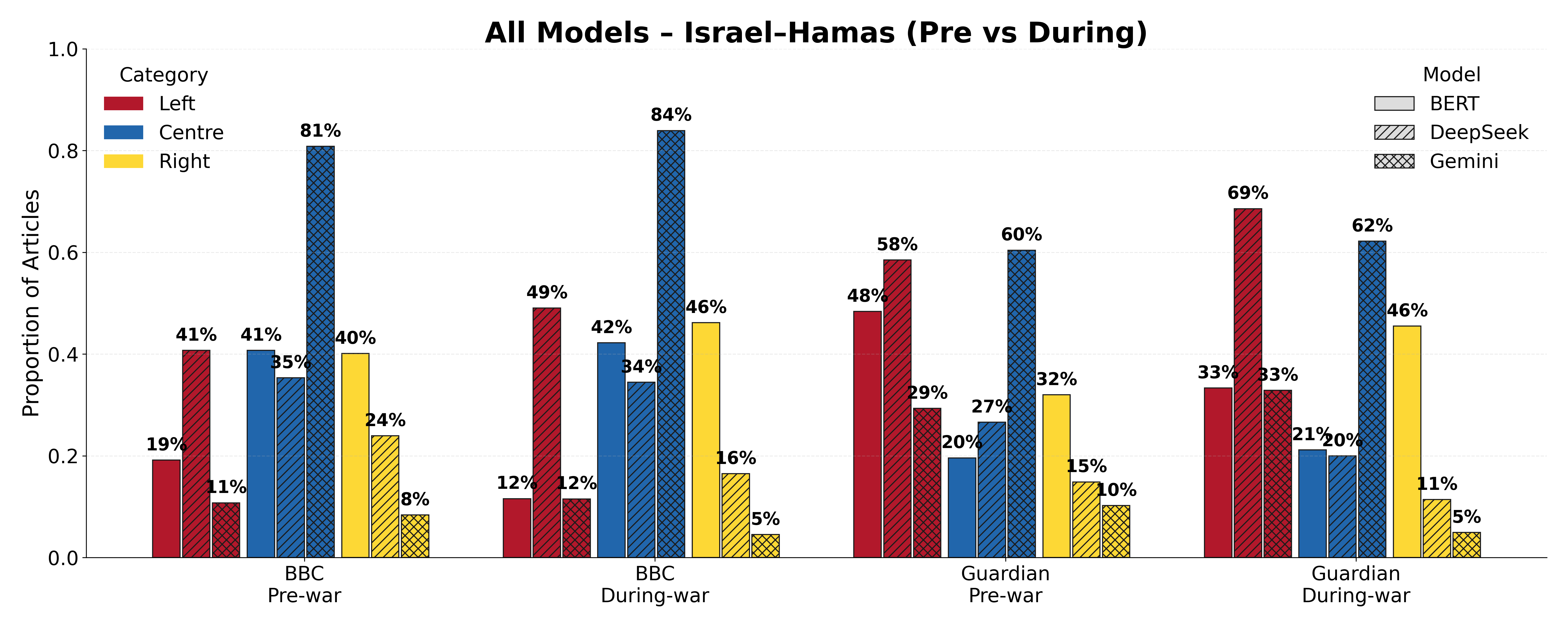}
  \caption{All Models – Israel–Hamas (Pre vs During). 
  Proportion of articles by Category (Left/Centre/Right) for BBC and \textit{The Guardian};
  hatch fills distinguish models (BERT, DeepSeek, Gemini).}
  \label{fig:all_models_ih}
\end{figure*}
 The implementation of political leaning analysis followed Steps 3-4 of our framework (Figure \ref{fig:workflow}). After the data cleaning and n-gram topic relevance validation processes, each news article was processed through the LLMs, where the models (BERT, Gemini 1.5 Flash, and DeepSeek-v3) assigned one of three labels - Left, Centre, or Right. In the case of BERT, we segment the input texts into overlapping 512-token windows with a stride of 256, and article-level labels were determined by majority voting across chunk-level predictions. Gemini and DeepSeek, which support long-article processing, analysed each article holistically via their APIs using a zero-temperature prompting approach to minimise randomness. We then aggregate the classified data results by temporal phase (pre-war vs during-war). These aggregated proportions form the basis for Figures 2–3 and Tables 4–5, enabling comparative evaluation of political leaning shifts across both time and models.

\subsubsection{Russia-Ukraine War}

In the Russia–Ukraine war  (conflict), we observe clear shifts in political orientation can be observed in the coverage of both the BBC and The Guardian before and during the two wars (Figure~\ref{fig:all_models_ru}, Table~\ref{tab:guardian_bias_summary} and Table~\ref{tab:bbc_bias_summary}). In the case of Bert, the BBC had a decline in left-leaning coverage, with the Centre becoming more prominent while the Right remained relatively stable. In the case of Gemini, during the war, the proportion of Neutral coverage in the BBC increased to 83.9\%, with a slight rise in Left to 11.5\% and a notable decline in Right to 4.5\%. 

BERT shows that the Guardian's trend before and during the war reflects a reduction in left-oriented reporting and a modest increase in right-leaning coverage, suggesting a rightward movement overall. In the case of Gemini, the share of Neutral rose to 62.2\%, Left increased to 32.9\%, and Right decreased to 4.9\%. Overall, after the outbreak of the conflict, both BBC and Guardian were assessed by Gemini as becoming more concentrated in the Neutral position, although the Guardian showed increases in Left-leaning but decreased in Right-leaning content alongside this Neutral shift.

In the case of DeepSeek (Figure~\ref{fig:all_models_ru}), pre-war BBC reports mainly focused on Centre leaning (64.0\%), and the  Guardian had  Left-wing and Centre leaning predominant. During the war, the BBC's left-wing proportion substantially rose to 33.0\%, central dropped to 53.7\%, right-wing slightly dropped to 13.2\%; the Guardian's central rose to 42.9\%, left-wing slightly dropped to 48.7\%, right-wing dropped to 8.4\%. 

Overall, during-war BBC under DeepSeek's judgment showed obvious increase in left-wing content, a decrease in central, while Guardian showed reports slightly more concentrated on central, accompanied by minor decrease in left-wing and right-wing, with BBC's left-wing rise most obvious (up 10.8\%), the largest change, central drop 10.3\%; Guardian's central up 1.6\%, right-wing drop 1.5\% larger. Data from Table~\ref{tab:guardian_bias_summary} and Table~\ref{tab:bbc_bias_summary} are consistent with chart results. DeepSeek values show BBC's Pre-war Left 22.2\% rising to During-war Left 33.0\%, while Pre-war Central from 64.0\% dropping to 53.7\%, further highlighting the trend of left-wing rise and central drop.




\subsubsection{Israel–Hamas War}

The results of the Israel–Hamas war, the BERT model shows the BBC recorded 19\% Left, 41\% Centre, and 40\% Right before the war, compared to 12\% Left, 42\% Centre, and 46\% Right during the war (Figure~\ref{fig:all_models_ih}, Table~\ref{tab:guardian_bias_summary} and Table~\ref{tab:bbc_bias_summary}). This demonstrates a significant decline in left-oriented reporting alongside a rise in both Centre and Right coverage, forming a more balanced but right-leaning profile. The Guardian showed a stronger pre-war left orientation (48\% Left, 20\% Centre, 32\% Right), but during the war this shifted to 33\% Left, 21\% Centre, and 46\% Right. Here again, the Left-leaning coverage decreased sharply, while Right-leaning content increased considerably.


In the case of Gemini, before the war, BBC coverage was almost entirely classified as Neutral (89.0\%), while the Guardian was primarily Neutral (79.8\%). During the war, the proportion of Neutral in the BBC remained stable at 89.3\%, with Left decreasing to 5.8\% and Right slightly increasing to 4.9\%. For the Guardian, the share of Neutral rose markedly to 87.0\%, while Left declined to 9.7\% and Right decreased to 3.3\%. Overall, after the outbreak of the conflict, the BBC  consistently maintained a highly Neutral reporting stance with almost no substantial changes, whereas the Guardian demonstrated a stronger shift toward Neutral, accompanied by reductions in both Left-leaning and Right-leaning content, with the decline in Left being more pronounced.





DeepSeek in Figure~\ref{fig:all_models_ih} shows that
pre-war,  BBC had Left-wing  Centre coverage  predominant (40.7\% and 35.3\%); Guardian had Left-wing  as the highest coverage (58.5\%). During the war, the BBC's Left-wing proportion substantially rose to 49.0\%, Centre slightly dropped to 34.5\%. Guardian's Centre-leaning coverage dropped to 20.0\%, left-wing rose to 68.6\%, and Right-wing dropped to 11.4\%. During the war, DeepSeek further shows that the BBC and the Guardian showed a trend of decreased Right-wing content, with the BBC's Left-wing increase the most (up 8.3\%), while the Guardian's Left-wing coverage rose to 10.1\%.  Tables~\ref{tab:guardian_bias_summary} and  ~\ref{tab:bbc_bias_summary} further confirm this trend. 

\begin{table}[!t]
\centering
\captionsetup{justification=centering}
\caption{Guardian – Political Bias Summary}
\label{tab:guardian_bias_summary}
\setlength{\tabcolsep}{3pt}
\footnotesize
\resizebox{0.9\columnwidth}{!}{
\begin{tabular}{lccc}
\toprule
\textbf{Metric} & \textbf{BERT} & \textbf{DeepSeek} & \textbf{Gemini} \\
\midrule
IP During-war Centre & 0.212 & 0.200 & 0.622 \\
IP During-war Left   & 0.333 & 0.686 & 0.329 \\
IP During-war Right  & 0.455 & 0.114 & 0.049 \\
IP Pre-war Centre    & 0.196 & 0.266 & 0.604 \\
IP Pre-war Left      & 0.484 & 0.585 & 0.293 \\
IP Pre-war Right     & 0.320 & 0.149 & 0.102 \\
RU During-war Centre & 0.240 & 0.429 & 0.870 \\
RU During-war Left   & 0.320 & 0.487 & 0.097 \\
RU During-war Right  & 0.440 & 0.084 & 0.033 \\
RU Pre-war Centre    & 0.226 & 0.413 & 0.798 \\
RU Pre-war Left      & 0.371 & 0.488 & 0.156 \\
RU Pre-war Right     & 0.403 & 0.099 & 0.046 \\
\bottomrule
\end{tabular}
}
\end{table}

\begin{table}[!t]
\centering
\captionsetup{justification=centering}
\caption{BBC – Political Bias Summary}
\label{tab:bbc_bias_summary}
\setlength{\tabcolsep}{3pt}
\footnotesize
\resizebox{0.9\columnwidth}{!}{
\begin{tabular}{lccc}
\toprule
\textbf{Event - Bias Type} & \textbf{BERT} & \textbf{DeepSeek} & \textbf{Gemini} \\
\midrule
IP During-war Centre & 0.422 & 0.345 & 0.839 \\
IP During-war Left   & 0.116 & 0.490 & 0.115 \\
IP During-war Right  & 0.462 & 0.165 & 0.045 \\
IP Pre-war Centre    & 0.407 & 0.353 & 0.808 \\
IP Pre-war Left      & 0.192 & 0.407 & 0.108 \\
IP Pre-war Right     & 0.401 & 0.240 & 0.084 \\
RU During-war Centre & 0.420 & 0.537 & 0.893 \\
RU During-war Left   & 0.160 & 0.330 & 0.058 \\
RU During-war Right  & 0.420 & 0.132 & 0.049 \\
RU Pre-war Centre    & 0.357 & 0.640 & 0.890 \\
RU Pre-war Left      & 0.193 & 0.222 & 0.069 \\
RU Pre-war Right     & 0.450 & 0.138 & 0.040 \\
\bottomrule
\end{tabular}
}
\end{table}

\begin{figure*}[!t]
    \centering
    \begin{subfigure}{0.49\textwidth}
        \centering
        \includegraphics[width=\linewidth]{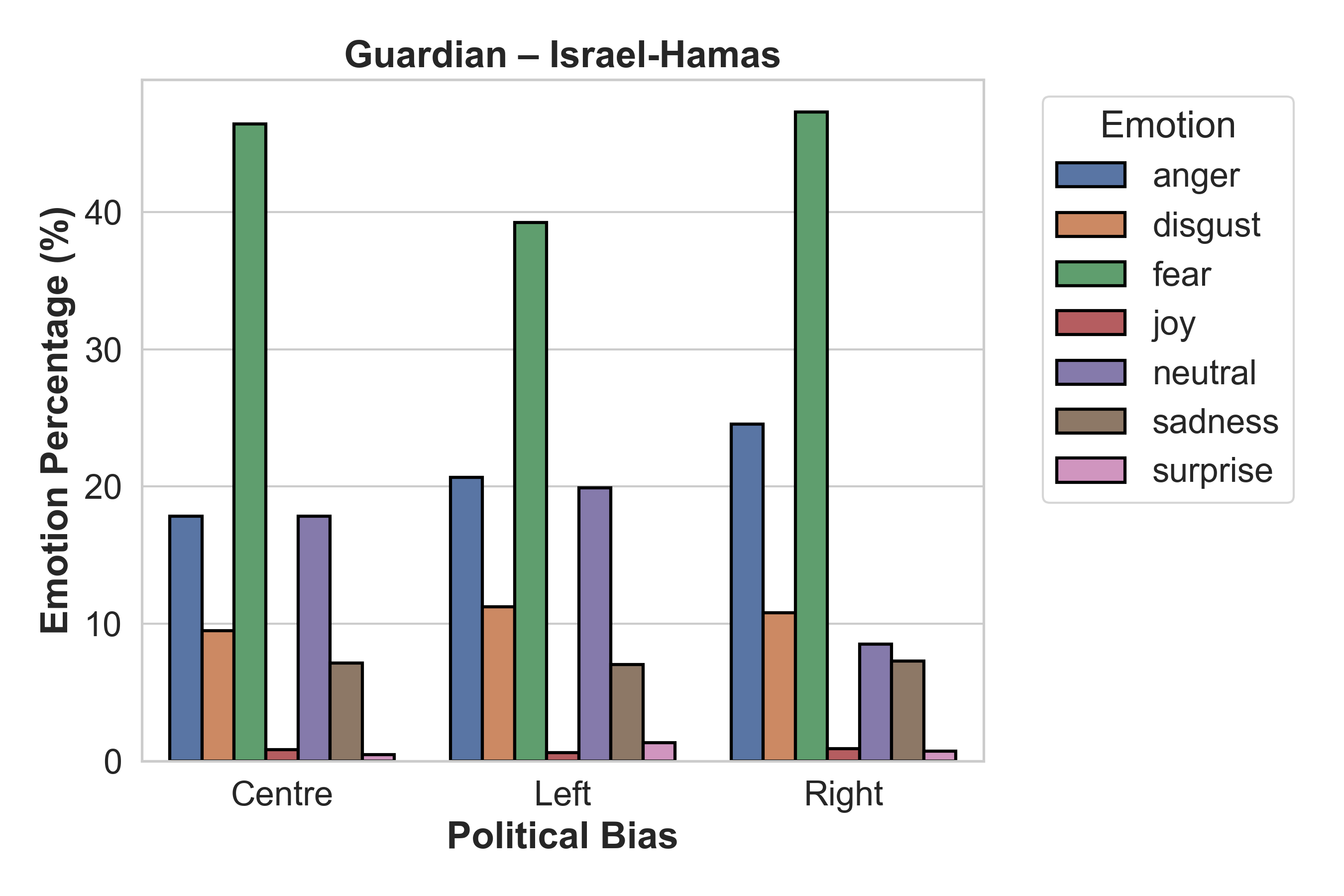}
        \caption{Guardian – Israel–Hamas}
    \end{subfigure}\hfill
    \begin{subfigure}{0.49\textwidth}
        \centering
        \includegraphics[width=\linewidth]{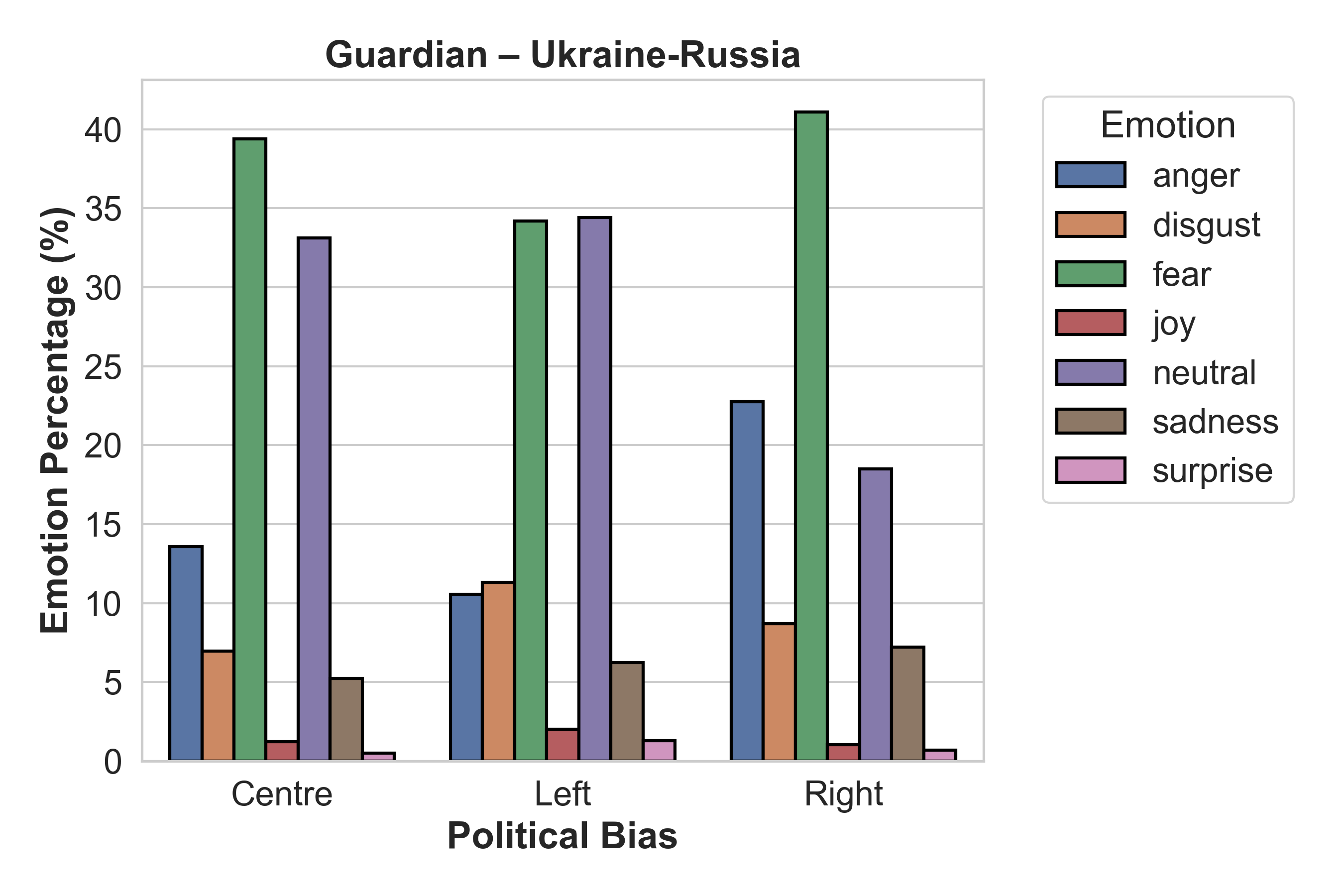}
        \caption{Guardian – Ukraine–Russia}
    \end{subfigure}

    \vspace{0.8em}

    \begin{subfigure}{0.49\textwidth}
        \centering
        \includegraphics[width=\linewidth]{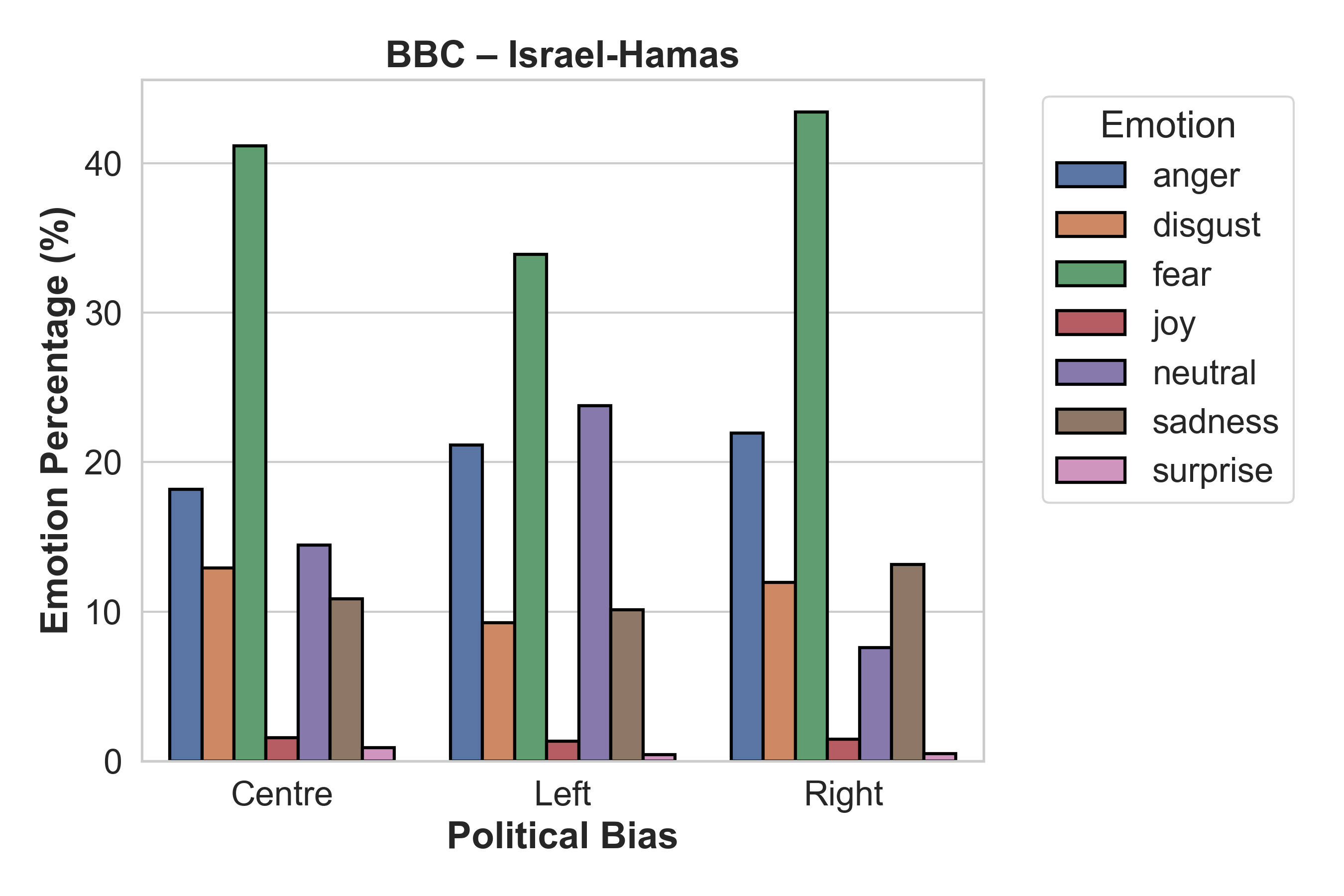}
        \caption{BBC – Israel–Hamas}
    \end{subfigure}\hfill
    \begin{subfigure}{0.49\textwidth}
        \centering
        \includegraphics[width=\linewidth]{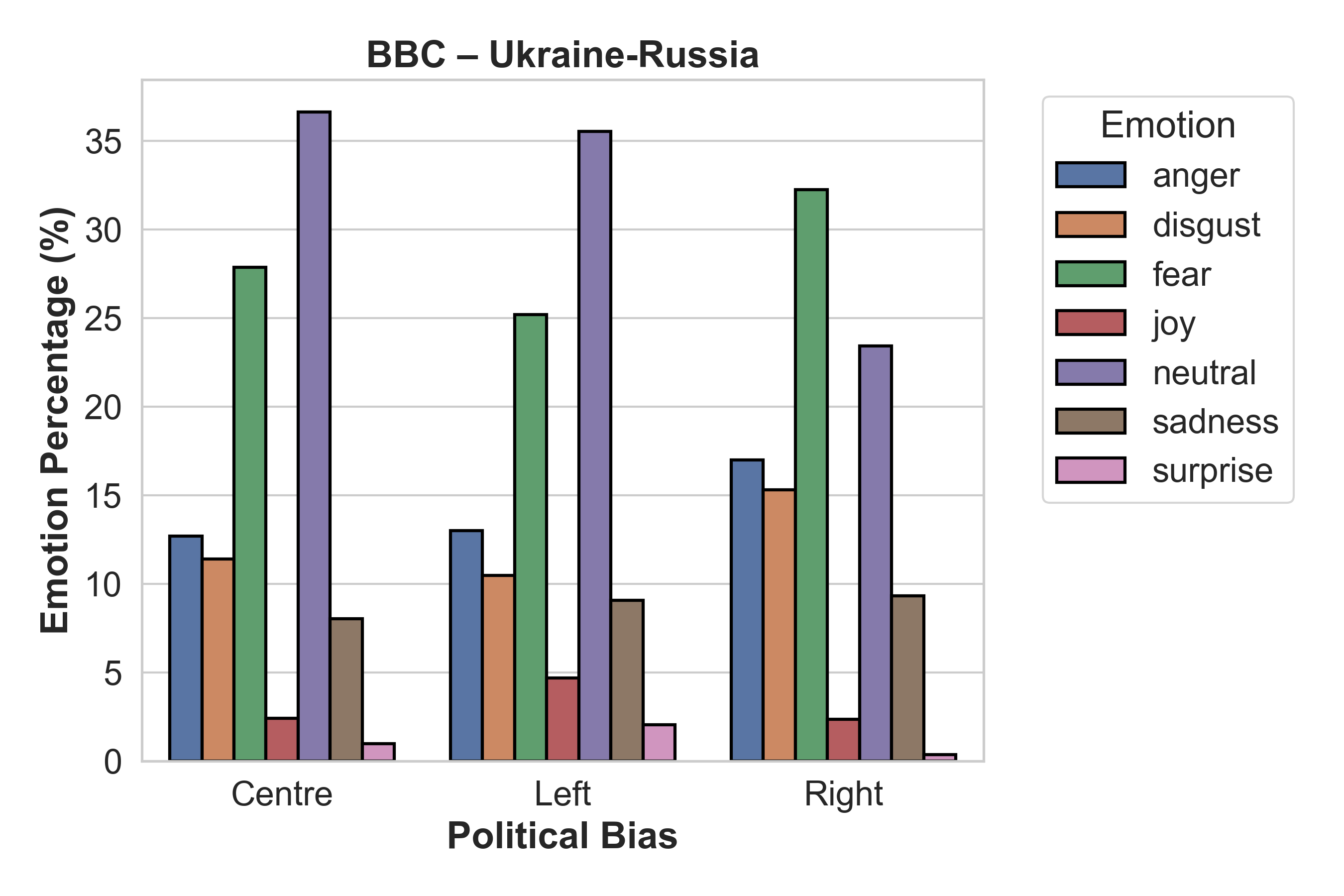}
        \caption{BBC – Ukraine–Russia}
    \end{subfigure}

    \captionsetup{justification=centering}
    \caption{The political leaning bias (score) variation across time for BBC and the Guardian under the Israel–Hamas and Russia–Ukraine conflicts.}
    \label{fig:media_conflict_trends}
\end{figure*}

\subsection{Sentiment Analysis}


We present results for sentiment analysis   to complement the political leaning classification by revealing the emotional tone underlying media discourse. We employed the pre-trained DistilRoBERTa model  to classify each article into one of seven emotion categories: anger, disgust, fear, joy, sadness, surprise, and neutral. We aggregate he sentiment predictions by media source (BBC and The Guardian), conflict (Russia–Ukraine and Israel–Hamas), and time period (pre-war vs during-war). These aggregated results were further cross-referenced with the political leaning labels to explore correlations between emotional expression and ideological stance.

Figure~\ref{fig:media_conflict_trends} displays the emotional distribution across news articles with different political leanings (Centre, Left, Right) for both BBC and \textit{The Guardian}, covering the Russia--Ukraine and Israel--Hamas conflicts. Figure~\ref{fig:media_conflict_trends}~(a) presents \textit{The Guardian}’s coverage of the Israel--Hamas war. The Fear sentiment dominates across all political leanings, particularly in Right-leaning articles (approximately 43\%). We observe that both anger and fear appear slightly more frequently in Left and Right-leaning news articles compared to Centre-leaning. The proportion of Neutral emotion is relatively higher in Left-leaning articles, suggesting a more restrained tone, whereas Right-leaning articles tend to be more emotionally charged.Figure~\ref{fig:media_conflict_trends}~~(b) shows \textit{The Guardian}’s coverage of the Russia--Ukraine war. Similar to the Israel--Hamas conflict, fear remains the leading emotion. However, neutral emotion significantly increases, accounting for more than 35\% in both centre- and left-leaning articles. Right-leaning articles exhibit elevated levels of anger and disgust, implying a more intense emotional tone.

Figure~\ref{fig:media_conflict_trends}~~(c) illustrates the BBC’s reporting on the Israel--Hamas conflict. Fear again dominates, especially in Right-leaning articles, where it reaches nearly 47\%. Anger is also notably higher in Right-leaning coverage (around 25\%). In contrast, articles with centrist and left-leaning bias exhibit more Neutral emotion, indicating a more measured tone. Figure~\ref{fig:media_conflict_trends}~~(d) depicts the BBC’s coverage of the Russia--Ukraine war, where Neutral and Fear are the two most prominent emotions. However, anger significantly rises in right-wing articles, while sadness and neutrality dominate in Left and Centre-leaning articles, which implies a stronger subjective stance and emotional intensity in right-leaning narratives.

\subsection{Political leaning bias  over time}

\begin{figure*}[!t]
  \centering
  \begin{subfigure}{0.49\textwidth}
    \centering
    \includegraphics[width=\linewidth]{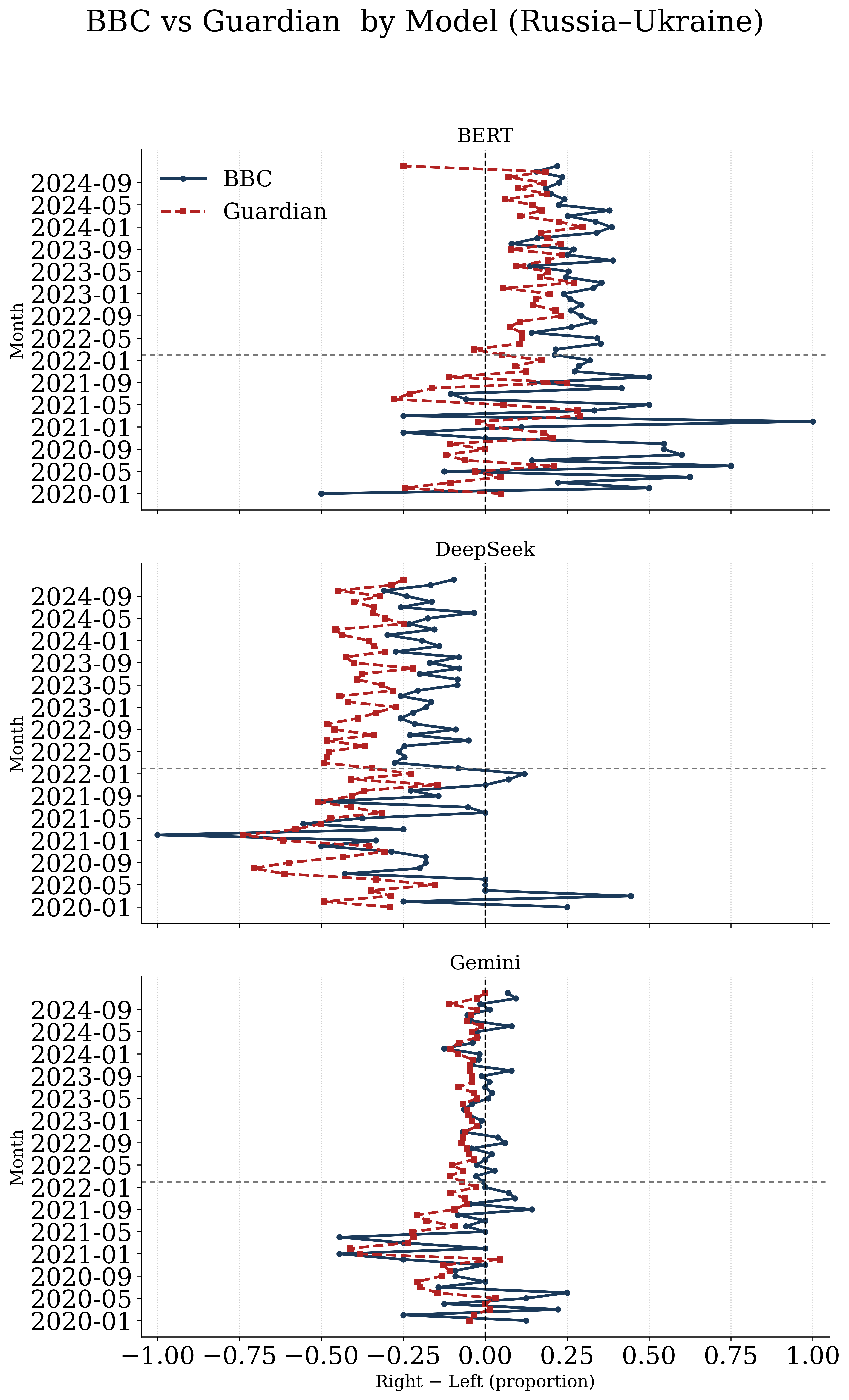}
    \caption{Russia--Ukraine (Right -- Left = Difference)}
    \label{fig:RLdiff_RU}
  \end{subfigure}
  \hfill
  \begin{subfigure}{0.49\textwidth}
    \centering
    \includegraphics[width=\linewidth]{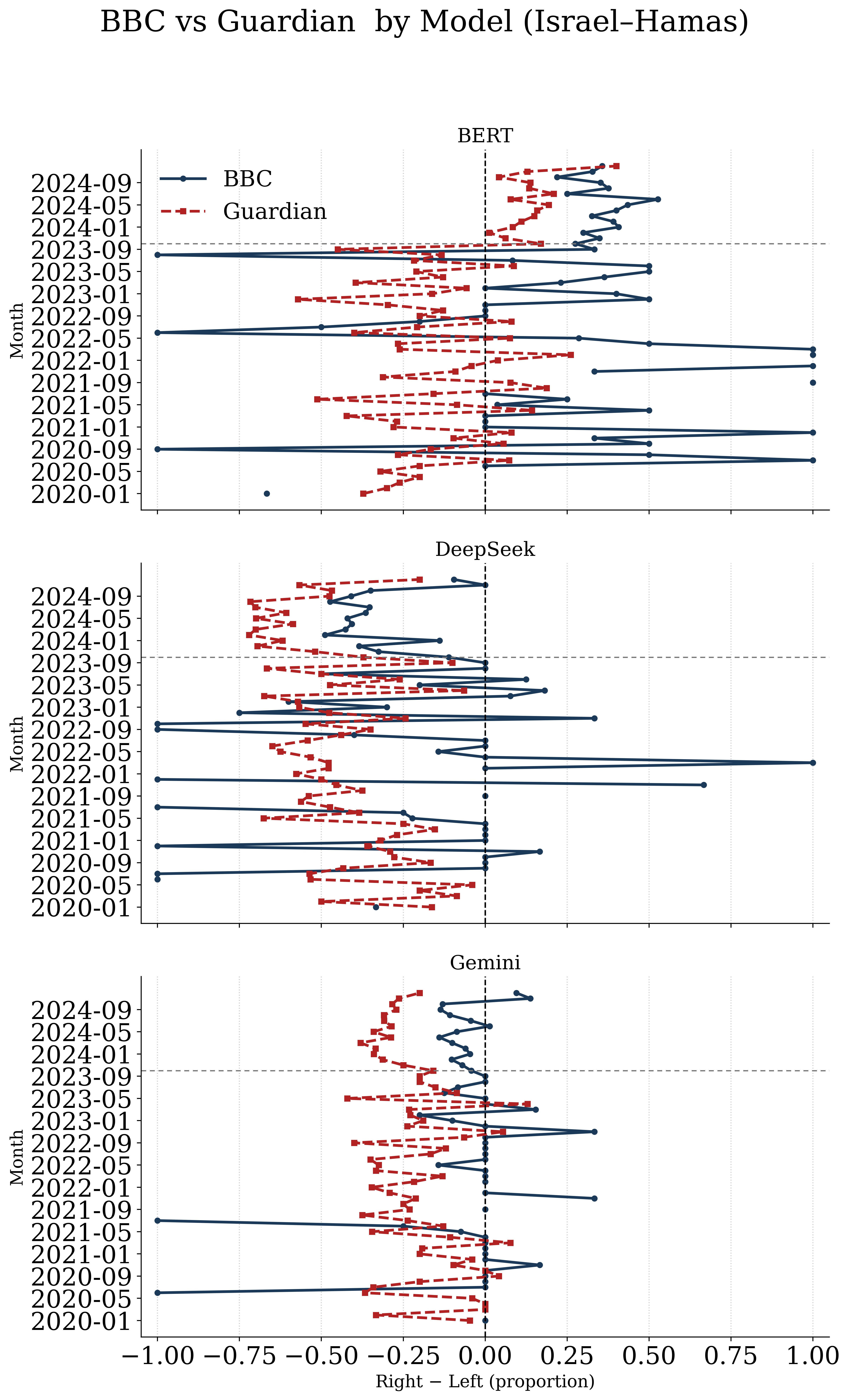}
    \caption{Israel--Hamas (Right -- Left = Difference)}
    \label{fig:RLdiff_IP}
  \end{subfigure}

  \captionsetup{justification=centering}
  \caption{Comparison of Left and Right difference between BBC and \textit{The Guardian} by three models (BERT, DeepSeek, Gemini). Each panel shows monthly trends across the two conflicts. Right -- Left = Difference, hence negative territory indicates Left-leaning, and positive territory indicates Right.}
  \label{fig:RLdiff_both}
\end{figure*}

We assess the change in political leaning (bias) before and after major geopolitical events and longitudinally compare how news agencies differ.Figure~\ref{fig:RLdiff_both} illustrates the difference between Left-leaning and Right-leaning coverage for BBC and the Guardian using BERT, DeepSeek, and Gemini. The Centre (neutral) leaning has been excluded to emphasise the direct contrast between Left and Right leanings. The positive values indicate a higher proportion of Right-leaning coverage, while the negative values indicate a higher proportion of Left-leaning coverage.

The outbreak of the war in February 2022 marks a turning point for the Russia–Ukraine war (Figure~\ref{fig:RLdiff_RU}). The pre-war coverage by both outlets fluctuated with higher volatility between Left and Right positions. However, after the outbreak, both the BBC and The Guardian exhibited narrower swings. A similar dynamic is observable in the Israel–Hamas context (Figure~\ref{fig:RLdiff_IP}), where October 2023 marks visible shifts in reporting. The major geopolitical event, such as the outbreak of a war, seems to correlate with reduced swing in political bias in news reporting.

\subsection{Comparison of language models}


\begin{figure*}[htb]
    \centering
   \includegraphics[width=\textwidth]{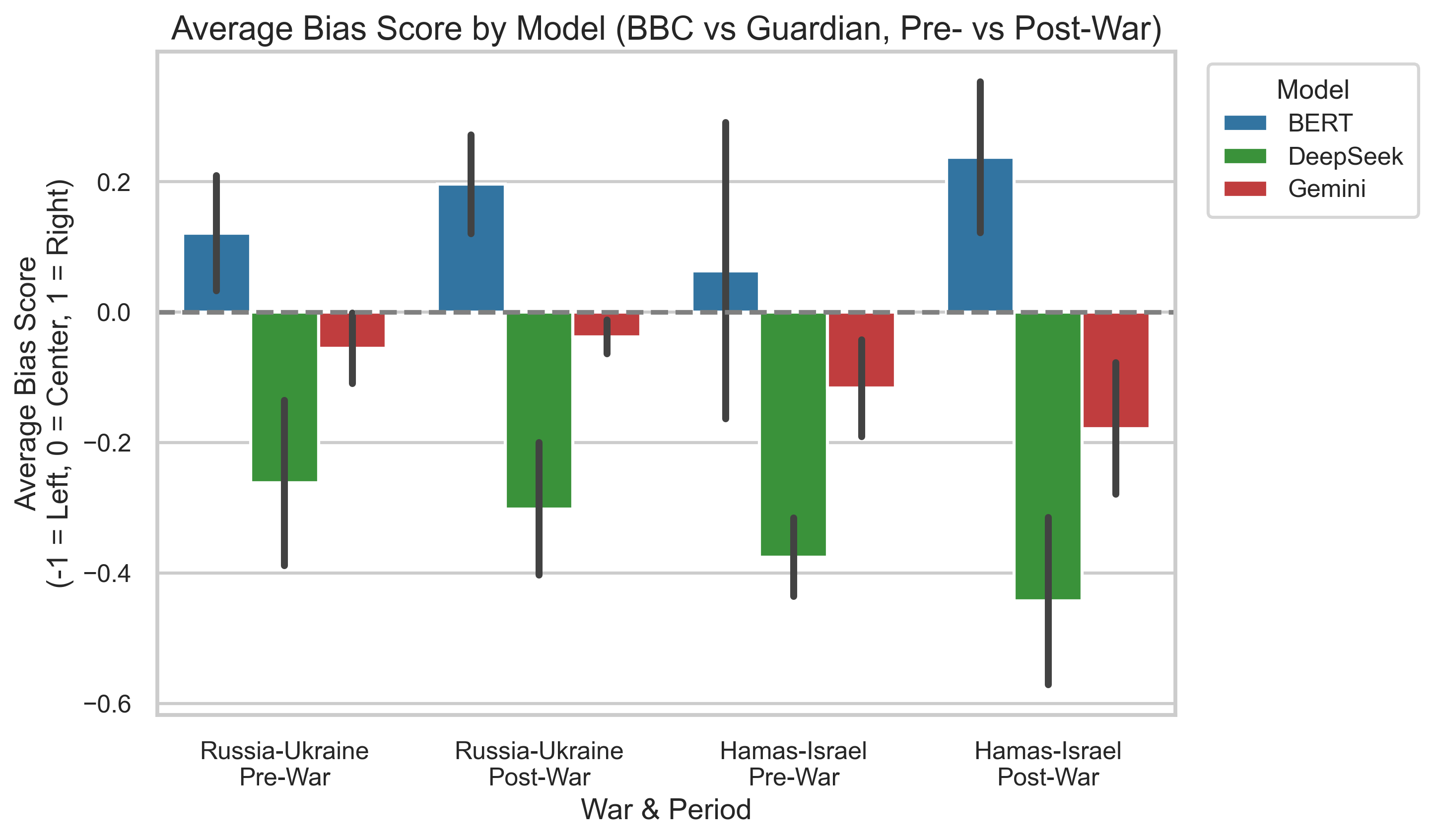} 
   \caption{Average bias score by model (BERT, DeepSeek, Gemini) during four war periods: Russia–Ukraine (pre/post) and Hamas–Israel (pre/post). A positive score indicates right-leaning bias, a negative score indicates left-leaning bias, and a score of zero represents a neutral position.}
    \label{fig:bias_score_barplot}
\end{figure*}

\begin{table}[H]
\centering
\caption{Average Political Bias Score by Model and War Period}
\label{tab:bias_scores}
\begin{tabular}{lccc}
\hline
\textbf{War \& Period} & \textbf{BERT} & \textbf{Deepseek} & \textbf{Gemini} \\
\hline
 Ukraine--Russia Pre-War   & 0.12 & -0.26 & -0.06 \\
 Ukraine--Russia During-War & 0.20 & -0.30 & -0.04 \\
 Israel--Hamas Pre-War    & 0.06 & -0.38 & -0.12 \\
 Israel--Hamas During-War & 0.24 & -0.44 & -0.18 \\
\hline
\end{tabular}
\end{table}

We next compare the average political bias scores predicted by the language models using  BBC and The Guardian. Figure~\ref{fig:bias_score_barplot} and Table~\ref{tab:bias_scores} show the average bias scores predicted by three language models (BERT, DeepSeek, and Gemini) for  the Russia–Ukraine    and the Hamas–Israel wars, during both pre-war and post-war periods. The vertical axis  in Figure~\ref{fig:bias_score_barplot} represents the bias score, where positive values indicate a right-leaning stance, negative values indicate a left-leaning stance, and zero suggests a centrist position. 
Noticeable differences can be observed in how each model assesses bias for the same data. For example, DeepSeek consistently produces negative scores across all scenarios, indicating a left-leaning classification. In contrast, BERT generally outputs positive scores, reflecting a right-leaning tendency. The scores from Gemini remain closer to zero, suggesting a more centrist evaluation. These differences are likely influenced by the models’ training data and internal classification standards.

We also observe distinct patterns when comparing media sources, where BERT shows a slight increase in right-leaning bias for BBC in the post-war period for the Russia–Ukraine war. DeepSeek indicates a stronger leftward shift, particularly for Guardian reporting after the war. Gemini shows minimal change across both media outlets and periods. In the case of the Hamas–Israel war, the divergence between models becomes more pronounced. BERT reports a notable increase in right-leaning bias for BBC post-war coverage. DeepSeek reveals further intensification of left-leaning scores, especially for Guardian articles, reaching the most negative value among all scenarios. Gemini continues to report relatively neutral or slightly left-leaning scores with little variation. Overall, BERT and DeepSeek show consistent right- and left-leaning tendencies, respectively, while Gemini presents more balanced outcomes. This variation across models highlights the uncertainty and divergence present in bias detection by large language models. It also suggests that model background and internal bias should be carefully considered when interpreting such results.


\vspace{0.5em}

\begin{table}[!t]
    \centering
    \captionsetup{justification=centering}
    \caption{Political  Bias Scores by Model, Media, and Conflict}
    \label{tab:radar_bias_scores}
    \setlength{\tabcolsep}{3pt} 
    \footnotesize 
    \resizebox{0.95\columnwidth}{!}{ 
    \begin{tabular}{l l l r}
        \toprule
        \textbf{Model} & \textbf{Media} & \textbf{Conflict} & \textbf{Bias Score} \\
        \midrule
        BERT      & BBC      & \textit{Israel--Hamas}      &  0.35 \\
        BERT      & BBC      & \textit{Russia--Ukraine}    &  0.21 \\
        BERT      & Guardian & \textit{Israel--Hamas}      &  0.12 \\
        BERT      & Guardian & \textit{Russia--Ukraine}    &  0.03 \\
        DeepSeek  & BBC      & \textit{Israel--Hamas}      & -0.31 \\
        DeepSeek  & BBC      & \textit{Russia--Ukraine}    & -0.14 \\
        DeepSeek  & Guardian & \textit{Israel--Hamas}      & -0.57 \\
        DeepSeek  & Guardian & \textit{Russia--Ukraine}    & -0.39 \\
        Gemini    & BBC      & \textit{Israel--Hamas}      & -0.08 \\
        Gemini    & BBC      & \textit{Russia--Ukraine}    & -0.00 \\
        Gemini    & Guardian & \textit{Israel--Hamas}      & -0.28 \\
        Gemini    & Guardian & \textit{Russia--Ukraine}    & -0.11 \\
        \bottomrule
    \end{tabular}
    }
\end{table}

\paragraph{Longitudinal analysis}

We constructed a weekly political bias index series for each language model to investigate how media bias predictions evolve across the Russia–Ukraine and Israel–Palestine wars. Each index represents the average predicted leaning of all articles published in the particular week, with bias labels numerically encoded  (Left = –1, Centre = 0, and Right = +1) with major geopolitical events annotated on the timeline to facilitate interpretation.

\begin{figure*}[!t]
    \centering
    \includegraphics[width=14cm, height=0.30\linewidth]{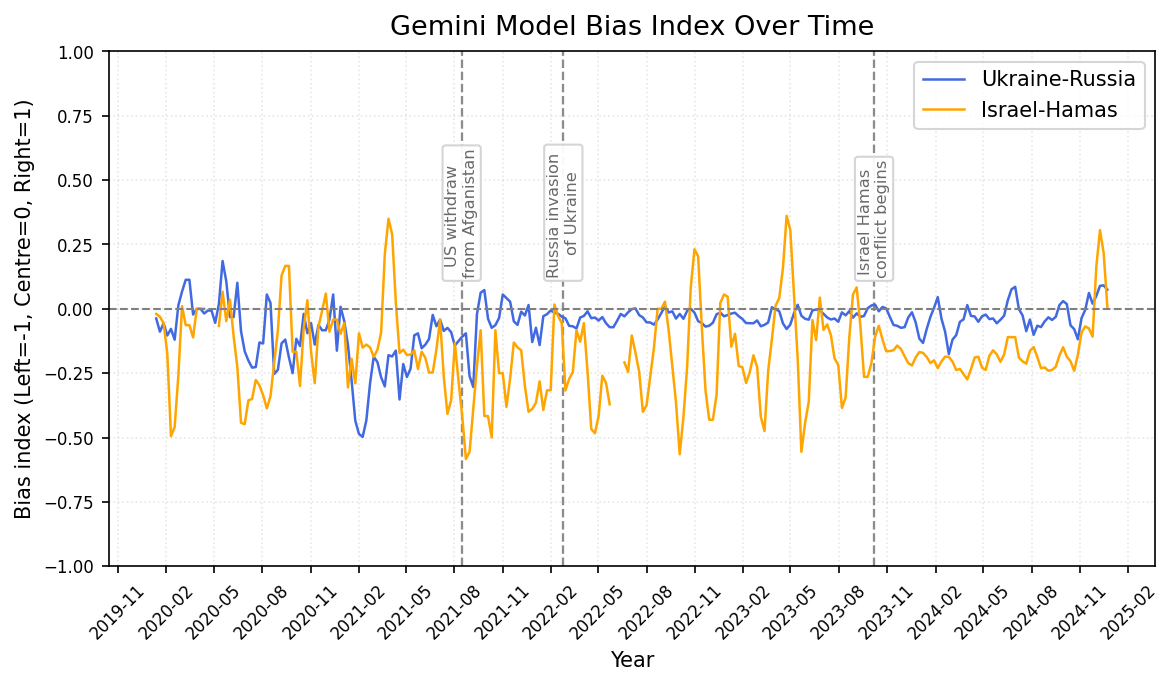}
    \vspace{0.6em}
    \includegraphics[width=14cm, height=0.30\linewidth]{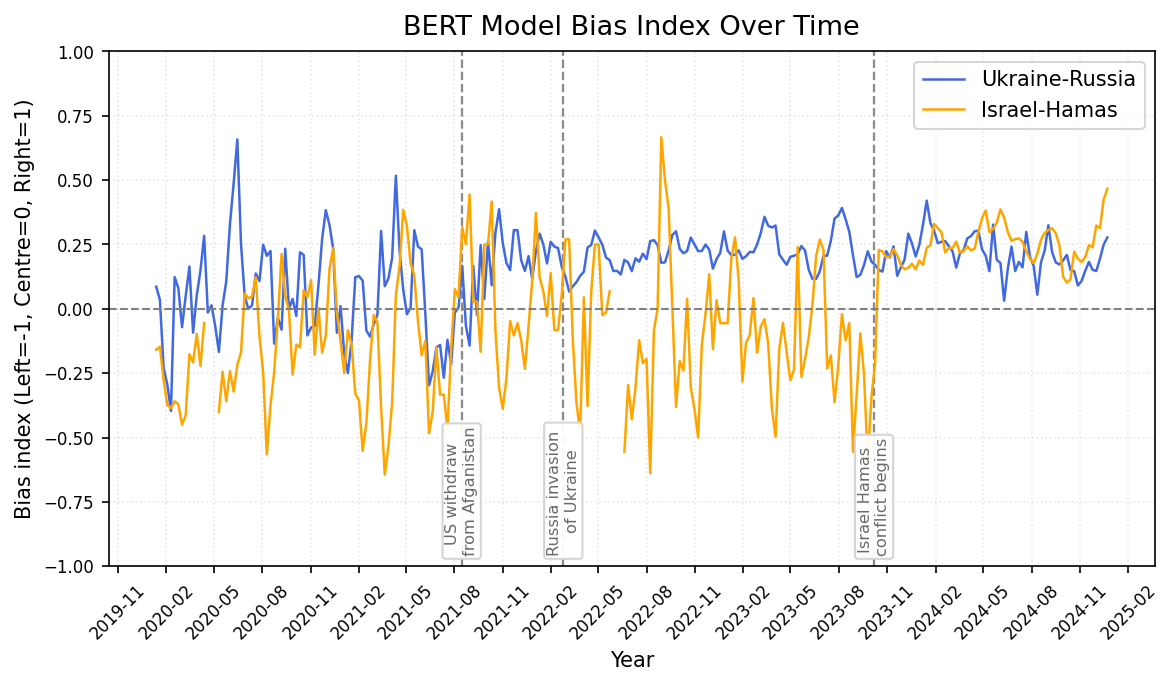}
    \vspace{0.6em}
    \includegraphics[width=14cm, height=0.30\linewidth]{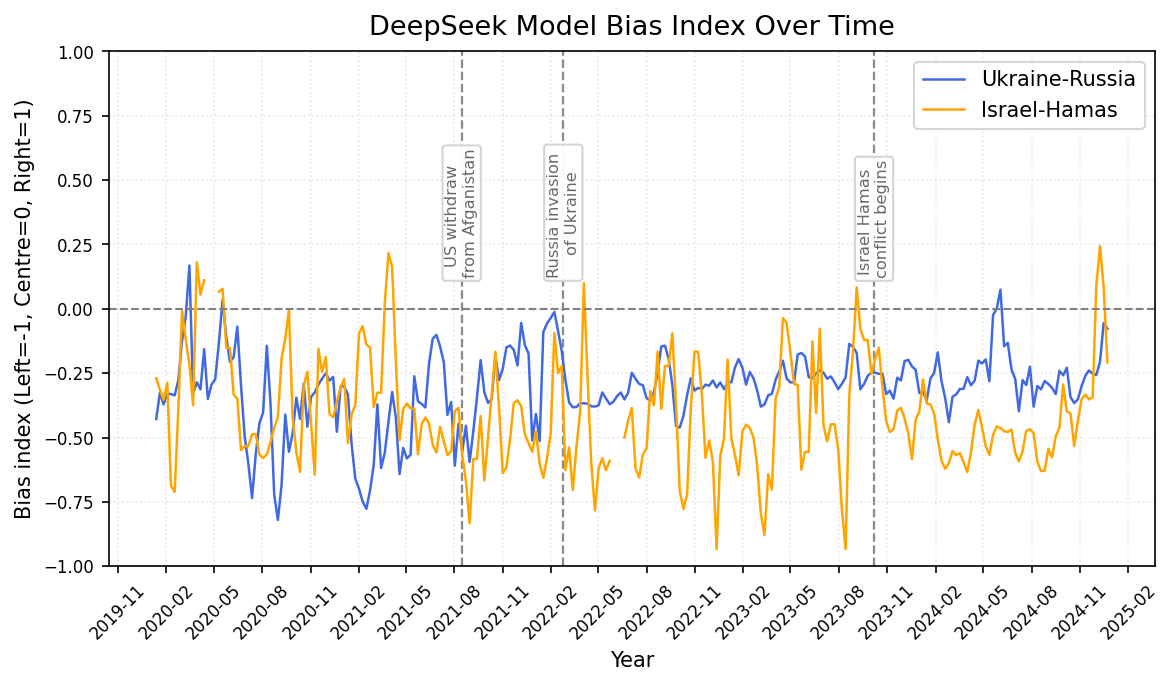}
    \captionsetup{justification=centering}
    \caption{Bias Index Over Time for Ukraine--Russia and Israel--Hamas conflicts across three large language models (Gemini, BERT, and DeepSeek). Major geopolitical events such as the US withdrawal from Afghanistan, Russia’s invasion of Ukraine, and the onset of the Israel--Hamas conflict are annotated.}
    \label{fig:llm_bias_trends_vertical}
\end{figure*}


Figure \ref{fig:llm_bias_trends_vertical} reveals how different language models perceive shifts in media leaning over time by capturing structural variations in coverage across key geopolitical moments, given distinct sensitivity levels.
We can observe that BERT displays strong reactivity to major events; eg, during the Oct 2023 Gaza conflict, its Israel--Hamas bias index surged from –1.0 (left-leaning) to +0.17 (right-leaning) within a week, suggesting a sudden perception of pro-Israel framing. Similarly, during the onset of the Russia--Ukraine war in February 2022, its Russia--Ukraine bias dropped from +0.24 to nearly neutral, indicating a shift from hawkish narratives to a broader anti-war consensus.
 We can observe that DeepSeek maintained consistent left-leaning predictions, especially around humanitarian crises. Although  DeepSeek did react to events such as the 2021 Gaza conflict or Russia--Ukraine invasion, the changes were smoother and more gradual. Notably, DeepSeek perceived the media as persistently critical of militarism and aligned with human rights narratives, particularly for the Israel--Hamas case.

 Gemini in Figure \ref{fig:llm_bias_trends_vertical}  remained the most stable across both conflicts and all time points; its bias index hovered near neutrality with minimal fluctuations. Even during intense crises, such as the 2022 Russia-Ukraine invasion or 2023 Hamas attacks, Gemini did not register a significant deviation. This suggests robustness to noise but limited responsiveness to real-time agenda changes.

In general, BERT demonstrated the most sensitivity to unfolding events, capable of identifying sharp turning points in media tone. DeepSeek followed the same directional trends but with a delayed and dampened response. Gemini served as a control baseline—its outputs stable but possibly under-responsive. These distinctions highlight how model architectures and internal reasoning affect perceived bias trajectories. For applications prioritising responsiveness to real-time sentiment (e.g., conflict monitoring), BERT may offer better analysis, and for historical or structural analysis, Gemini’s stability would be an advantage.

\section{Discussion}


 In this study, we systematically examined how  LLMs can be used for longitudinal study of  political bias in Western media. Our analysis addressed three interrelated dimensions: (i) how political leaning within the same outlet shifts before and after major geopolitical events, (ii) how BBC and The Guardian differ when reporting on the same conflict within a common time window, and (iii) how multiple LLMs (Political-BERT, DeepSeek-V3, and Gemini~1.5~Flash)  detect  bias when applied to identical news media content. Overall, the results show that geopolitical events induce measurable changes in the distribution of political labels; however, these shifts are mediated by outlet-specific editorial baselines and model-specific decision boundaries, rather than constituting a single universal ground truth of ``biased'' or ``unbiased'' coverage.


Across both conflicts, the two outlets (BBC and The Guardian) demonstrated distinct editorial tendencies but similar movement tendencies (Figure 5). BBC's overall coverage is observably closer to the neutral position but marked by larger fluctuations, sometimes leaning rightward and other times leftward depending on the month, especially in the pre-war periods. The Guardian is consistently positioned on the left-leaning side across models, with fewer instances of rightward position (Table 8). This pattern holds across both conflicts, suggesting that while geopolitical events amplify volatility in both outlets, the BBC tends to oscillate between ideological poles, whereas The Guardian sustains a stable left orientation. Notably, despite these differences, both outlets often move in the same direction in response to major events, indicating a tendency for correlated shifts in bias over time.
These results suggest a degree of outlet-specific consistency across conflicts. Regardless of whether the context was Russia–Ukraine or Israel–Hamas, BBC coverage exhibited greater variability and responsiveness to external shocks, while The Guardian maintained a persistent leftward leaning (Table 8). This reinforces the notion that political bias in media is shaped not only by global events but also by the intrinsic editorial stance of the outlet. At the same time, both outlets often moved in parallel directions in response to major events, reinforcing the notion that political bias in media is shaped not only by global events but also by the intrinsic editorial stance of the outlet.


 These findings should be interpreted in light of several limitations. First, our analysis is restricted to two UK-based, English-language outlets and two highly salient conflicts within a fixed window, so the findings cannot be generalised to other regions, languages and cultural groups. Second, we constructed the datasets  using keyword and tag filtering plus web scraping, which may exclude relevant articles or include only marginally related pieces, and they inherit any editorial choices embedded in platform metadata. Third, we map each article to a single left/centre/right label and then aggregate at the monthly or phase level, which simplifies within-article nuance, mixed framing, and section-level differences. Fourth, our models (Political-BERT, DeepSeek-V3, and Gemini~1.5~Flash) reflect biases based their   training data and internal definitions of ideological categories; disagreements between models highlight that our labels are model-dependent proxies rather than a gold-standard ground truth. Finally, this study does not incorporate systematic human annotation for calibration, so our assessment of correctness and bias remains indirect. This motivates future work that combines LLM predictions with expert-coded benchmarks to better audit both media outlets and the models used to evaluate them.

Future studies could retain the structured text while supplementing inputs with raw audio and video, leveraging the long-context capacity and native audio/video pipelines of multimodal LLMs to achieve finer-grained evidence chain tracking, event timeline reconstruction, and cross-modal consistency validation. This would enable a more comprehensive and temporally detailed portrayal of media narratives and political bias~\cite{google2025keyword_audio,ai2025_models,ai2025_longcontext,videomme2024,googlecloud2025_expand}. Furthermore, given access to other media APIs   (such as \textit{The New York Times}, Reuters, or Al Jazeera),  it would be possible to compare media coverage across a wider range of cultural and political contexts. This would allow for a more comprehensive examination of the differentiated characteristics of international public opinion in conflict reporting~\cite{reese2001framing,pierri2022propaganda} for major historical/political events.

In this study, we structured the news text before feeding it into the model, which helps ensure field consistency, reproducibility, and robustness in cross-model comparisons. However, with the rapid evolution of multimodal LLMs, it may  be feasible in the future to process unstructured data such as audio and video directly. This would reduce the need for cumbersome preprocessing and transcription while preserving more contextual cues and temporal semantics. For example, Gemini 2.5 natively supports real-time understanding and conversational interaction with streaming audio and video, can perform real-time analysis based on screen sharing or video streams, and is capable of contextual awareness regarding when to respond or ignore background sounds~\cite{google2025keyword_audio,ai2025_models}. At the same time, the Gemini series offers a context window on the scale of millions of tokens, enabling complex retrieval and reasoning across documents and modalities~\cite{ai2025_longcontext}. From a third-party evaluation perspective, video benchmarks targeting long-term and cross-modal understanding (such as Video-MME) have already demonstrated significant progress in the performance of commercial multimodal models on video question-answering and comprehension tasks, providing a useful reference for incorporating video material directly into future analyses~\cite{videomme2024}. In terms of practical deployment, enterprise platforms are also expanding the capabilities of Gemini 2.5 Pro/Flash to support lower-latency voice/video interaction and complex agent-based workflows, which means that in production environments, we can more reliably integrate audio and video sources to build an end-to-end multimodal news analysis pipeline~\cite{googlecloud2025_expand}.

\section{Conclusions}

We developed a political bias detection framework based on three language models (BERT, DeepSeek, and Gemini) to systematically analyse shifts in media stance within BBC and The Guardian during two major geopolitical events: the Russia-Ukraine war and the Israel-Hamas war.  

We observed a clear systematic difference between models, where  DeepSeek consistently showed a stable Left-leaning tendency, while BERT and Gemini remained closer to the Centre. These differences appeared not only in absolute bias scores but also in how trends were interpreted across different scenarios. Our results demonstrate that the BBC and The Guardian showed distinct reporting behaviours across the two conflicts. In the Russia-Ukraine war, both outlets maintained relatively stable positions, and the models detected only minor bias shifts. However, in the Israel--Hamas conflict, we identified larger political bias  shifts particularly in Guardian coverage, suggesting a more event-driven pattern of reporting bias. DeepSeek was more sensitive to shifts in media position and better captured changes in reporting direction before and after conflict outbreaks and Gemini  showed a comparatively   conservative bias strategy. These variations suggest that LLMs are shaped not only by their training data and architecture, but also by underlying "worldviews" and classification standards.


Our study revealed distint patterns of political bias in mainstream media reporting during major international conflicts. We highlighted the strengths and weaknesses of selected LLMs in political bias detection tasks. Future work could explore model perceptions and reliability using broader datasets and a human-in-the-loop for annotation and qualitative assessment in our framework. 

\section*{Data and Code Availability}
    We provide open source code and data that can be used to extend this study \footnote{\url{https://github.com/sydney-machine-learning/politicalbias-LLMs}}.

    
\appendix
 \bibliographystyle{elsarticle-num} 
 \bibliography{cas-refs}





\end{document}